\documentclass[manuscript]{aastex}
\usepackage{outlines}
\usepackage{enumitem}

\slugcomment{to appear in The Astrophysical Journal}

\shorttitle{Kepler Starspots}
\shortauthors{Basri}

\begin{document}


\title{Calibration of Differential Light Curves \break 
for Physical Analysis of Starspots}


\author{Gibor Basri}
\affil{Astronomy Department, University of California,
    Berkeley, CA 94720}

\email{gbbasri@berkeley.edu} 





\begin{abstract}

This paper presents detailed consideration of methodologies to calibrate differential light curves for accurate physical starspot modeling. We use the Sun and starspot models as a testbed to highlight some factors in this calibration that that have not yet been treated with care. One unambiguously successful procedure for converting a differential light curve into a light deficit curve appears difficult to implement, but methodologies are presented that work in many cases. The years-long time coverage of {\it Kepler} provides a strong advantage, but unresolved issues concerning the competing and sometimes similar effects of surface differential rotation versus spot number and size evolution can prevent the confident recovery of correct spot covering fractions in certain cases. 

We also consider whether faculae are detected by {\it Kepler} and/or must be accounted for. We conclude their effects are such that absolute photometry is not required for spot deficit calibrations. To elucidate their signature we re-examine correlations between absolute brightness, differential variability, and apparent spot coverage for hundreds of {\it Kepler} stars with absolute calibrations from \citet{Mont17}. The results are similar to theirs, but we draw somewhat different conclusions. Most of the stars in this active solar-type sample are spot-dominated as expected. Partly because of a dearth of longer period stars, the evidence for facular dominance in this sample is both sparse and relatively weak. The facular population exhibits a puzzling lack of dependence on rotation period, which raises questions about the apparent detection of a ``facular" signal at short periods.

\end{abstract}





\section{Introduction\label{sec:intro}}

Sunspots are relatively dark regions in the solar photosphere caused by inhibited local convection due to concentrated magnetic fields. They have been observed over centuries, yielding insights into the Sun's surface differential rotation and the operation of its magnetic dynamo. Photometric changes in other stars ascribed to starspots have also been observed for more than a century \citep{Stras09}, but with vastly poorer time coverage and precision and little detailed information on their sizes and positions. There is good evidence that on very young and/or rapidly rotating stars, the spot patterns live for many rotations, with a tendency to be concentrated towards polar regions, whereas they appear at mid to low latitudes on the Sun and only survive for at most a couple of rotations. Starspot coverage on younger stars can often be far larger than for sunspots, producing much greater amplitudes of photometric variability \citep{Bas11}.

With the advent of precision space photometry, particularly the COROT \citep{Bag03} and Kepler \citep{Bor10} space telescopes, the observational situation has been markedly improved. Both missions produced light curves continuously over at least a few months with very fine time resolution for many thousands of stars. These provide very precise relative differential changes in brightness on short time scales (hours, days), but may miss secular changes in absolute brightness (months, years). In examining these differential light curves, one of the most obvious features is that they often contain segments with systematic changes of amplitude over time. Many of these take the form of dips and peaks whose frequency is related to the stellar rotation period. Other than some cases of very rapid variations (over less than a couple of days) due to stellar pulsations, these are generally interpreted as diagnostic of starspots. In principle they encode information about stellar rotation periods and detailed information about stellar differential surface rotation, starspot coverage trends in both location and lifetime, possible active longitudes and polar spots, and stellar magnetic cycles \citet{Sav13}. There are, however, complications in actually using this data to infer starspot properties. Foremost among these is the fact that starspots produce a deficit relative to the unspotted star, but differential light curves are not the same as light deficit curves.

\citet{Moss09} provide an early example of some of the methods examined in this paper. They applied faculae-free analytic spot modeling to COROT light curves. They took the signal in the COROT Level-2 pipeline to be representative of relative stellar fluxes, except for a polynomial correction for detector drifts (similar to current practices with Kepler data). Thus, the brightest point in the differential light curve (which typically lasted over 2 months) was taken as the best estimate of the unspotted continuum, although they were explicitly aware of the fact that this is questionable and inhibits what can be deduced. They utilized a parameter search to find the best fitting models with fewer than 10 spots, and inferred rotation period, spot lifetime, and differential rotation for a few stars. 

Although the work in this paper reveals underlying uncertainties in their methods, their work is an example of what one hopes will eventually be able to be done more confidently. Other authors following somewhat similar procedures include \citet{Lanz09} and subsequent work by that group. One difference in modeling techniques is whether one uses a few large spots to fit the light curve (more recently with MCMC techniques), or divides the star into hundreds of pixels and uses a Maximum Entropy method to fit the light curve. These methodological questions are not the subject of this paper. Instead it is concerned with whether the light curves are calibrated in a way that allows one to be confident in the physical parameters that result from whatever fitting method one prefers.

The Kepler mission provides nearly continuous coverage over 4 years for well over a hundred thousand stars, with sufficient precision to detect individual sunspot groups if the Sun were hundreds of parsecs away. There is a great deal of information in these light curves, and many papers have been written using them. Those papers have wisely concentrated on general morphological properties of the light curves, most often the inference of rotation periods from them. Methods for doing that have been summarized by \citet{Aig15}, and fall under the general categories of using either periodograms or autocorrelation functions. In addition, a number of authors have tried to infer surface differential rotation from periodograms, including \citet{RRB13, Niel15, Das16, Sant17}. \citet{Ferr15} and \citet{Rein17} interpret long-term changes in amplitude as evidence of cyclic starspot coverage (activity cycles). The main issues discussed in this paper are not critical so long as one is only analyzing the morphology of the light curves (as in those studies) and not trying to deduce physical starspot properties. 

There have also been several explorations of simple starspot models, for example \citet{Niel12, RR13, Rein15, Sant17}. Other papers have tried to infer trends in spot lifetimes, for example \citet{Gil17, Ark18}. For reasons related to the motivation of this paper, there have been rather few investigations that specifically try to derive physical starspot properties from Kepler light curves. \citet{Ion16} model the star Kepler-210 with models of the sort we prefer here (a few large spots) and note that their model could be adjusted for spot size or temperature issues if the calibration of the light curve isn't right. The spatial resolution of broadband light curves is actually very poor (a significant fraction of a hemisphere), so the use of the term ``spot" should generally be understood as meaning ``regional spot distribution". 

\citet{Ion16} assume that differences in the phase of dips are due to differential rotation; we will argue below that this is risky since spot evolution can cause similar effects. The papers by \citet{Dav15, Sav16} are examples of analyses of one or a few Kepler stars in which a mixture of periodogram and spot modeling techniques are applied.  \citet{Bon12} also look at one star, but include a facular contribution. So far there has been little explicit evidence that this is necessary, although one would like to learn more about stellar faculae. In these and similar papers the differential light curve is taken at more or less face value, meaning that the non-magnetic star is almost always assumed to be represented by the brightest part. In this paper we give these assumptions a more critical examination.

    \section{What is the Problem?\label{sec:problem}}
    
There are characteristics of Kepler reduced light curves that should be understood before the data are used to infer properties of spots on stars, or produce starspot models. The primary one is that the data represent differential photometry that is only (partially) calibrated per quarter of observations. Starspots cause cause a deficit in flux compared to what an unspotted star would produce; it is this deficit that must be analyzed to infer starspot properties. One can transform a differential light curve into a deficit curve by setting the unspotted stellar flux to unity (the ``continuum") and interpreting the differential flux changes as deviations below that level, but two issues must first be resolved. 

One is whether the brightest point in the differential light curve should actually be set to unity (which assumes that at that moment the star is unspotted). This is only really justifiable if this brightest level occurs for (at least short) continuous segments of time. In fact there are almost never flat-topped bright peaks in the actual data for stars showing distinctive starspot dips. Thus all flux deficits must be understood to be lower limits to the true spot deficits, because any level of spottedness that persists throughout the observations is not accounted for. This problem is worse the shorter the observational period is, although even observations that last for years may suffer from it. For example, any persistent spottedness in the latitudes that are always visible due to stellar inclination is undetected, and the lower limit itself can evolve over time if that coverage changes. The second issue is whether there are calibration problems which mean that secular trends during a quarter due to either instrumental effects or changes in the overall stellar brightness have been removed by the reduction scheme. Such trends should be compensated for so that the unspotted star is properly represented by a constant value of unity.

Another problem inherent in this form of data is that because it is differential in nature and restricted in time, it does not easily convey the effect of faculae. These are magnetic features that are generally found not only in the vicinity of starspots but in magnetically active regions that don't produce spots (e.g. \citet{Kos16} and references therein). An excellent summary of the effects of magnetic regions on precision photometry in the solar case has been presented by \citet{Shap17}. Facular regions have two opposite behaviors compared with starspots: they are brighter than non-magnetic regions of the star and they are more prominent near the stellar limb. This means their effects can be stronger at different rotational phases than spot effects, and inclination affects them differently from spots. Because of their more diffuse and extensive geometrical distribution, they tend to simply raise the general brightness of the signal rather than produce distinctive peaks analogous to the dips that starspots cause. Given the lack of absolute photometry in the Kepler pipeline light curves, the effects of faculae are typically not obvious. This issue is examined in more detail below. For all the reasons above, the pipeline light curves are not suitable for physical starspot analysis without further manipulation. This paper is partly an exploration of what those manipulations should be. 

\begin{figure}
    \begin{center}
    \epsscale{1.0}
    \plotone{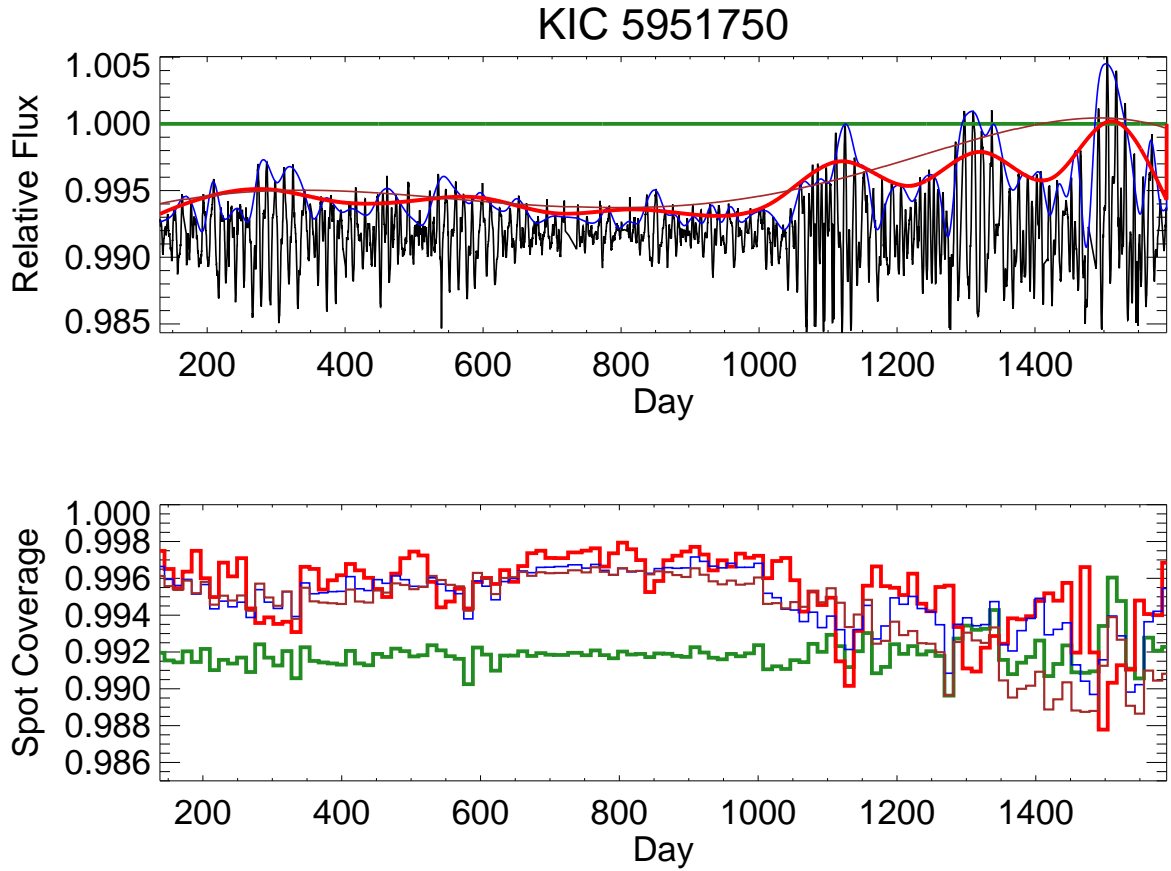}
    \caption{Various methods for fitting a ``continuum" to a differential light curve to enable its conversion to a light deficit curve suitable for inferring starspot coverage. The black curve in the upper panel is a normalized pipeline Kepler differential light curve. The different methods described in the text produce unspotted continuum fits (upper panel colored curves) to the pipeline curve; these can be used to produce light deficit curves. The Flat, CoarseFit, MidFit, and FineFit methods are shown using green, brown, red, and blue colors respectively. The lower panel shows the integrated spot coverage measured from the light deficit curves obtained using the methods in the upper panel. 
    \label{fig:Prob}}
    \end{center}
\end{figure}        

To illustrate the basic problem, consider the black curve shown in the upper panel of Fig. \ref{fig:Prob}, which is the full pipeline differential light curve of KIC 5951750. Suppose one only had the observation over a given 200 days. The easiest thing to do is to simply assume the unspotted continuum is represented by the brightest point in that observation, and divide by that value to normalize the light curve to unity. The peak-dip amplitude before about Day 1050 is significantly smaller than afterwards. Because observational time boundaries are not connected to anything stellar, the inferred deficit due to the dips near Day 950 would be considered much larger if they were measured using a normalization that included day 1150 (something like the green line) than if the observation terminated at day 1000 (something like the brown line). This is clearly unacceptable if one wants the correct measurement of the light deficit! Nonetheless, this is what most authors have explicitly done by taking the brightest point in their observations as the unspotted continuum. This problem exists whenever the observations do not contain times at which the star is actually unspotted. 

This problem does not arise when a transiting planet actually occults a starspot. Methods of studying the properties of spots under those circumstances (for example in \citet{Sil10, San11, Valio17, Morris17}) are much better constrained both in spot position and physical characteristics because the planet localizes it and the spot is compared to the local stellar continuum. When done over many rotations this also provides a more robust method for measuring differential rotation (at least at the latitudes sampled). Of course in the transit case there are also many spots present that are not occulted by the planet, so this method does not work for inferring global spot distributions or behavior. This paper confines itself to the inference of starspot properties from purely stellar light curves.

It is safer to use a light curve that lasts for several years to set the continuum (there is a greater chance of seeing the unspotted star), but this does not avoid all problems. The upper panel of Fig. \ref{fig:Prob} shows the result of using several different methods for assigning the unspotted continuum. The green line shows the continuum that results from the same (brightest point) method as above; we call this the ``Flat" method. The unspotted continuum is actually set to the value of the 99th percentile brightest point in the whole curve since Kepler data may contain flares, or noise that isn't stellar. It is important to keep in mind that the pipeline light curve (black line) has already had each quarter's median set to the same value, and there secular corrections have been applied within each quarter that tend to leave the median flat on timescales of a month or so.  

The deficit curve that results from using this continuum (normalizing it to unity) is shown as the green line in the middle panel. Notice that most of the light curve shows a substantial deficit relative to it because of the influence of the bright points near Day 1500, when the star was most variable. ``Variability" will generally be considered as a local quantity: the full amplitude of the differential light curve within a few rotations. Because the rotation periods of all the stars considered are known, the deficit over each whole rotation can be evaluated, yielding what we call the ``coverage". This is a measure of the integrated deficit due to spots over the whole star (except the parts permanently hidden by stellar inclination). We will refer to the coverage as ``greater" when the integrated deficit is larger; this corresponds to smaller values (further below unity) in the coverage plot, that indicate more spottedness. The coverage curve for the Flat case is the green line in the lower panel of Fig. \ref{fig:Prob}.

Because the level of coverage in the Flat case is strongly influenced by the most variable segment (which sets the continuum), it is similar whether the local variability is large or small. The one short excursion to less coverage just past Day 1500 is due to the fact that both the dips and peaks are higher there in the pipeline light curve. A general effect of using the Flat method to set the continuum is that the inferred coverage is relatively larger and constant throughout the (4 year) observation. This implies that the changes in the differential amplitude of the light curve over long timescales are mostly due to changes in the geometrical distribution of starspots rather than their total coverage. That is what would happen if the total spotted area does not evolve much, but differential rotation redistributes spots so that sometimes they are mostly on one hemisphere (producing high differential variability) and sometimes more evenly spread around (producing low variability). It could also happen if spots come and go (spot evolution) but the total spotted area is approximately constant. Either is possible, but unlikely to be true in the general case. Another issue is that if the observation stopped at Day 1000 then the inferred spot coverage would all be much lower. Of course, this is partly what was meant by the previous caution that a differential light curve provides only a lower limit to the true coverage. Exceptions to even this statement are discussed below.

If one does not trust the Flat method, then a different method of setting the unspotted continuum must be used. Fig. \ref{fig:Prob} illustrates three methods with differing degrees of time variability. The ``CoarseFit" method (brown curve in the upper panel) fits a third order polynomial to the 90th percentile bright points in time bins that are one quarter long (90 days). The ``FineFit" (blue curve) method uses a 3-point smoothed spline fit to the tops of each rotational peak. The intermediate ``MidFit" method also uses a spline fit (to the 75th percentile points), but the time bins are 8 rotation periods wide (this choice is justified in Section \ref{sec:Models}).

In all cases the pipeline curve is divided by the fit, then renormalized with the new 99th percentile point to produce an unspotted continuum at unity. The inferred integrated spot coverage can be found by integrating the deficit curves that result from these methods over each rotation period; these are shown in the lower panel of Fig. \ref{fig:Prob}. There are clear differences between the various methods, depending on the behavior of the light curve. In the region of highest variability the FineFit and MidFit curves agree that the coverage is generally less than from the Flat method, while the CoarseFit curve produces mostly greater coverage there. The discrepancies depend somewhat on the separation of regions of high versus moderate variability. In the region of lower variability all the fitting methods mostly agree with each other and produce much less coverage than the Flat method (particularly between days 600-1000).  

The task is to figure out which of these methods is closest to the truth under what circumstances. For that, an independent assessment of spot coverage is required. One approach would be to examine spectra at times of strong variability and weak variability. If molecular signatures of starspots \citep{Berd02} were detected, one could test whether they remained of equal strength (as the Flat method suggests) or were weaker when the variability is weaker (as the fitting methods suggest). There are contemporaneous spectra of some Kepler stars (particularly the planet-bearing ones), so this is an avenue that should be pursued. Another approach is to utilize absolute photometry to see whether the star is indeed darker when the coverage is found to be greater. This has the confounding problem of whether faculae actually tilt things in the other direction; the Sun is known to be absolutely brighter when it has more spots compared with when it has few (although this has nuances, as described in Section \ref{sec:Sun} below). 

Solar-type stars are generally known to fall into two classes when absolute brightness is compared with other measures of magnetic activity \citep{Hall09}: those which get darker with more activity (spot-dominated) and those which get brighter (faculae-dominated). A nice analysis of when this switch occurs (and why) can be found in \citet{Shap14}. Fortunately, recent work by \citet{Mont17} (hereafter MTF17) has produced absolute photometry for a few thousand solar-type Kepler stars, and they have also studied this question. We employ this additional information below to see if it helps with evaluating the methods of continuum fitting, and also revisit the brightness-variability analysis of MTF17 in Section \ref{sec:bvcorrl}. Before doing that, however, there are a couple of other ways to learn something relevant about the efficacy of the proposed calibration methods.

\subsection{The Sun as a Testbed\label{sec:Sun}}

It is always helpful to study a case for which one knows the answer in attempting to assess the efficacy of analysis techniques. In the case of the Sun, we have both precision photometry that measures the effects of sunspots on the total visible solar output, and contemporaneous imaging which shows the actual sunspot areal coverage generating the photometric deficit. The aim of this section is to transform the solar photometric signal into something approximating what Kepler would measure on it, and then to compare methods of converting that differential flux curve back to a deficit curve, and thus to see which method comes closest to yielding the best spot coverage. We also examine the effects of faculae by utilizing the absolute solar flux curve in the same way we will utilize the Kepler absolute fluxes from MTF17.

\begin{figure}
    \begin{center}
    \epsscale{1.0}
    \plotone{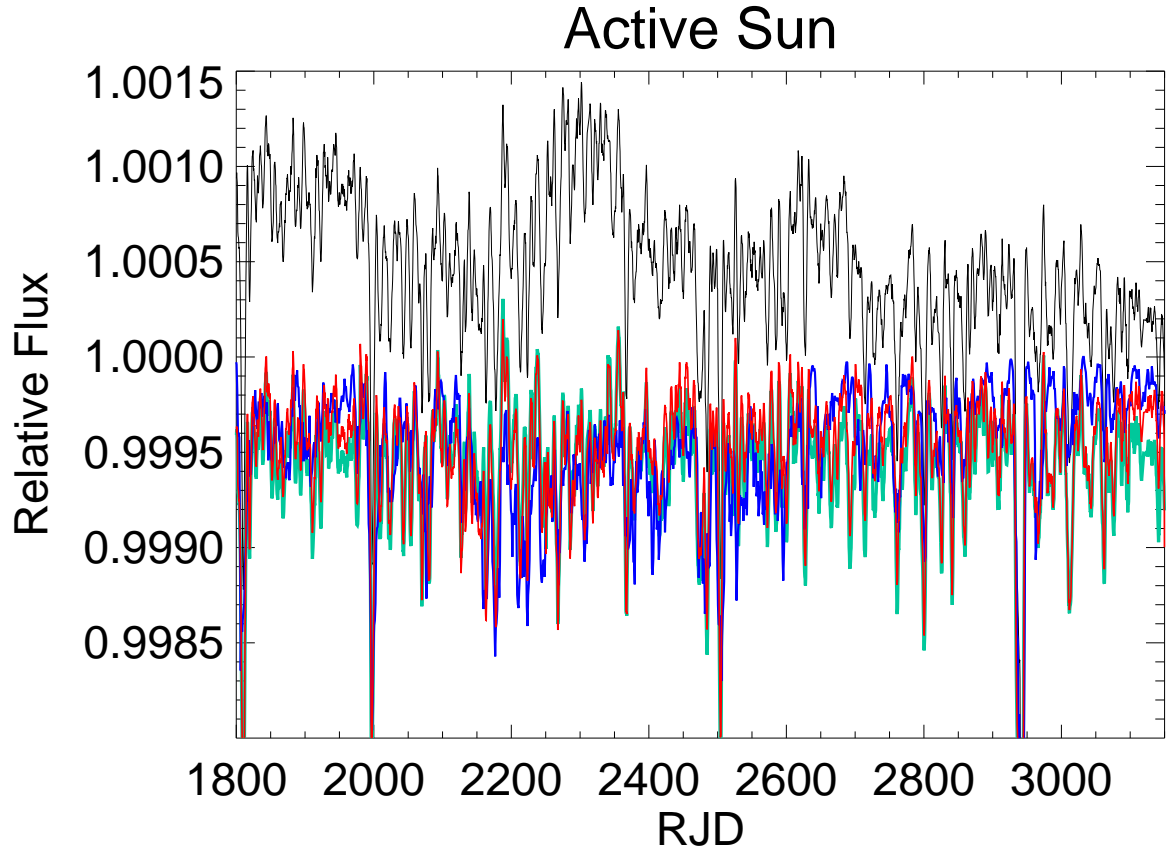}
    \caption{Active Sun. The black curve shows the absolute brightness (with a normalization of unity for the median of the absolute flux throughout Cycle 23). The green curve shows the result of converting this to a Kepler-like observation of 15 quarters (as described in the text) and then applying the Flat method. The red curve shows the same but using the MidFit method. The blue line shows a representation of the actual spot area as measured from solar images, normalized to a similar scale (and flipped over) to resemble the deficit curves.
    \label{fig:ActSun}}
    \end{center}
\end{figure}

\begin{figure}
    \begin{center}
    \epsscale{1.0}
    \plotone{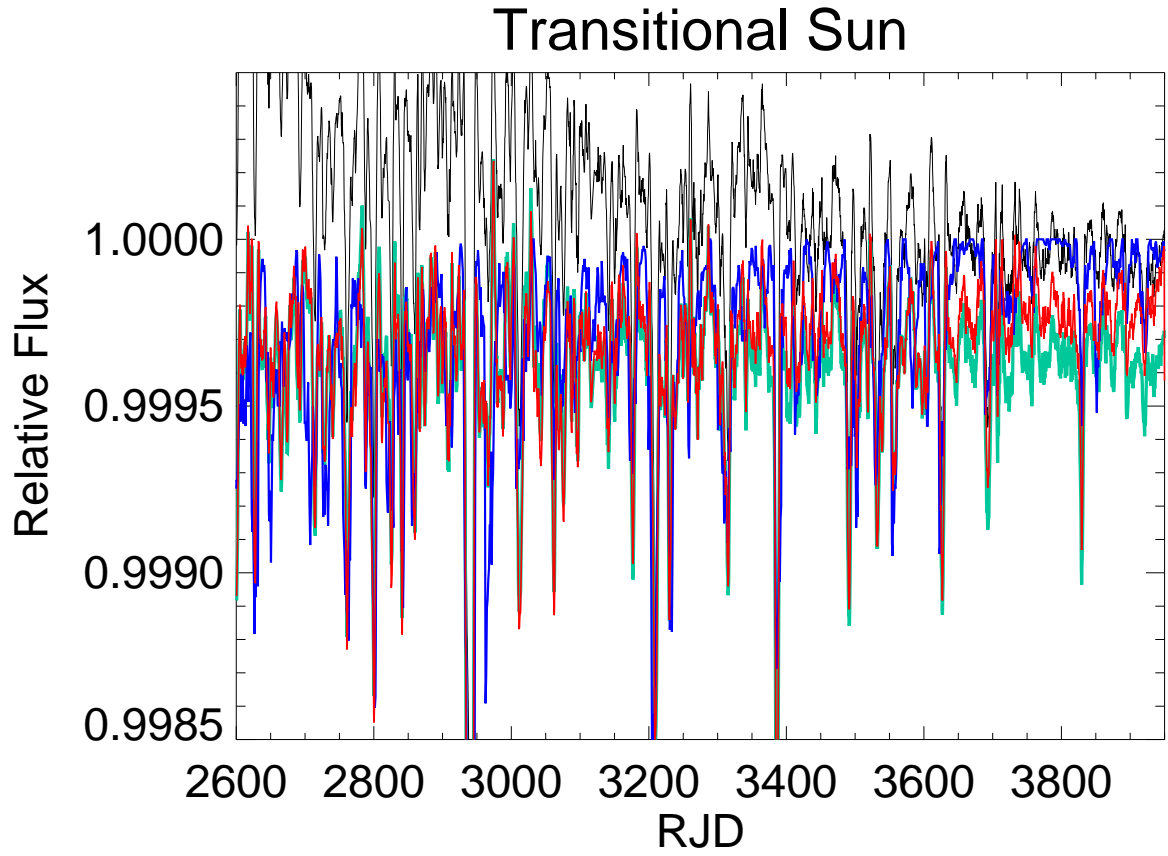}
    \caption{Transitional Sun. The same as Fig. \ref{fig:ActSun} but for a period during which the Sun is transitioning from active to quiet.
    \label{fig:TranSun}}
    \end{center}
\end{figure}

\begin{figure}
    \begin{center}
    \epsscale{1.0}
    \plotone{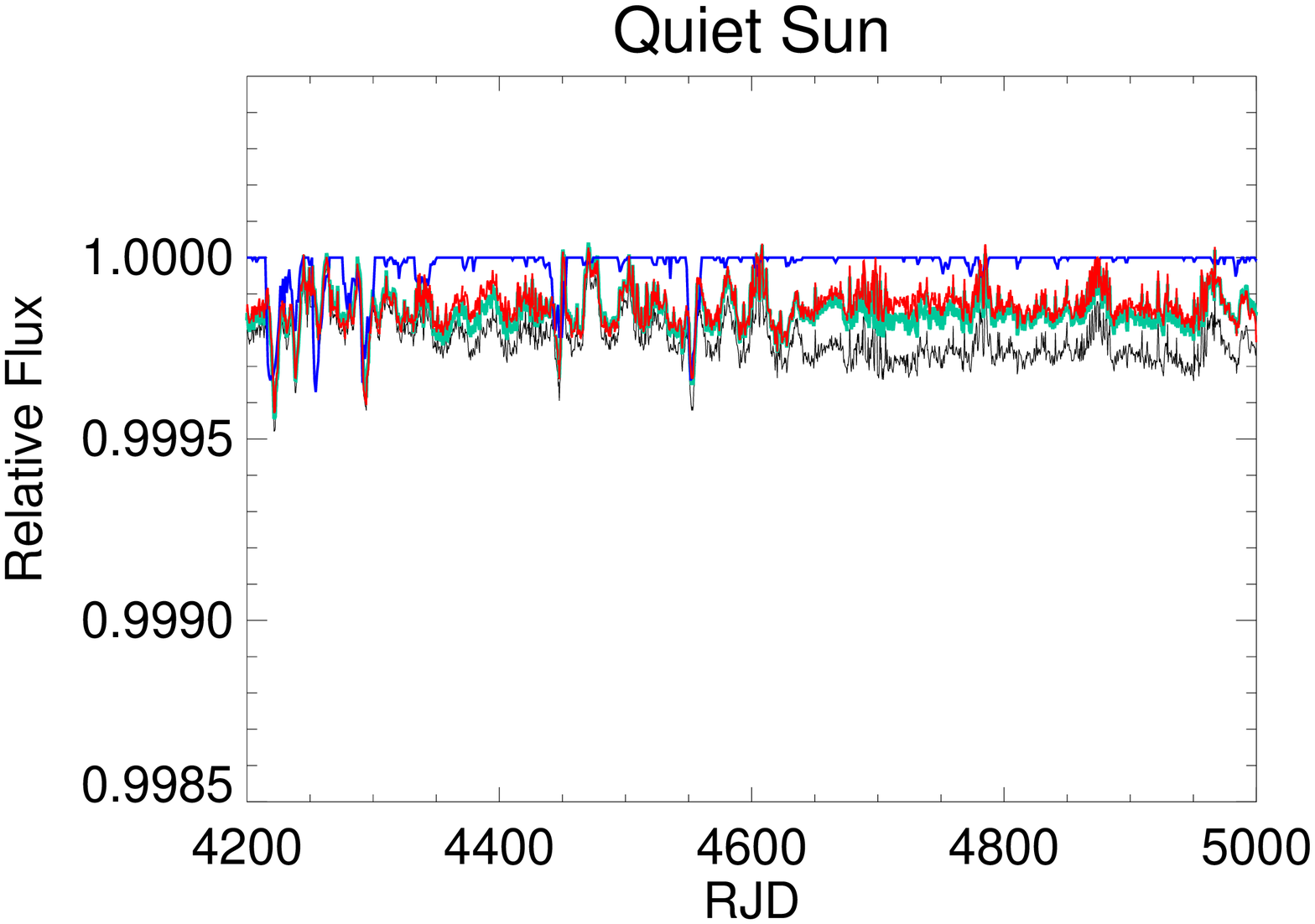}
    \caption{Quiet Sun. The same as Fig. \ref{fig:ActSun} but for a period during which the Sun is quiet (for 10 quarters). Here the Sun is actually fainter than its median cycle value, and there are significant periods during which no spots were seen (flat parts of the blue curve). The deficit curves are not completely flat due to the remaining facular signal.
    \label{fig:QuietSun}}
    \end{center}
\end{figure}

We begin with the same Virgo solar light curves from the SOHO spacecraft that were employed in past papers \citep{Bas13}. The absolute fluxes for all of Cycle 23 are normalized to a median of unity, then three time intervals are selected: one while the Sun was active (RJD 1800-3150), one while it was transitioning from active to quiet (RJD 2600-3950), and one while it was quiet (RJD 4000-5000). These curves are shown in black in Figs. \ref{fig:ActSun}-\ref{fig:QuietSun}. To produce Kepler-like light curves, each of these intervals is divided into 90 day segments then renormalized to unity and concatenated back together. The Flat (green) and MidFit (red) methods are then applied to produce light deficit curves. For the solar case (unlike the previous stellar example) there is not much difference between the fitting methods. Thus the CoarseFit and FineFit cases are not shown; these methods are dropped altogether for other reasons in Section \ref{sec:Models}. The reason is that the solar light curves don't have the very strong differences in variability that are present in the stellar example. The stellar sample in this paper is largely more rapidly rotating and active than the Sun, mostly because it is easier to determine rotation periods for them.

The observed umbral areas (originally in nano-hemispheres) from \citet{Hath17} are normalized, scaled, and flipped so that they resemble the light deficit curves (the scaling after normalization is division by a factor of 1.7 for the active Sun and 1.3 for the other two segments). In general, the deficit curves do a reasonable job of following the relative values of the underlying spot coverage (shown in blue) for the active and transitional Sun. There are clearly times when spots appear without causing the expected size of flux dip, and cases when a flux dip occurs without an indication of the concomitant spot coverage, but these are isolated instances. The MidFit solution is slightly superior in the quieter part of the transitional Sun, which also has the most contrast in variability. As before there is a suggestion that the Flat method tends to imply too much coverage in the less variable segments. This provides a weak case for preferring the MidFit method for stars, but is hardly definitive. 

A different problem is apparent in the quiet Sun, when there are significant periods during which the Sun is unspotted. Faculae still cause the solar brightness to vary, and the true solar rotation is more obvious in this part of the light curve without spots of varying sizes and growth/decay rates to confuse it. The Sun is fainter than average (recall that the average is over the whole magnetic cycle) despite the lack of spots, presumably because the facular contribution is reduced from the average. Indeed, the immaculate Sun (with no magnetic fields at all, so not the real Sun) might well be at an even slightly lower level. The methodology for converting flux to deficit curves makes the assumption that the brightest parts of the light curve are associated with the least spots; that is not actually true in the quiet Sun. It therefore sets the unspotted continuum at the peaks of the facular oscillations, implying an erroneous (though small) light deficit for most of the time that does not correspond with the actual spot coverage. This serves as a warning when interpreting Kepler light curves that have very low variability levels using this methodology. Nonetheless, these tests with the Sun generally affirm our approach to producing deficit curves from Kepler-like data.

In anticipation of the brightness-variability analysis pursued with Kepler stars in the Section \ref{sec:bvcorrl}, we now analyze the correlation of absolute brightness to spot variability for the Sun. The Sun is called an ``anti-correlated" star, meaning that it actually gets brighter in absolute terms when it is covered with more spots \citep{Shap14}. We have the full record of absolute brightness for the Sun. The Kepler pipeline products have no information on absolute brightness (other than the Kepler magnitude), but every month or so the Kepler spacecraft took a ``full-frame" image (FFI), associated with interruptions in the data stream for download to Earth or when rotating the spacecraft to the next quarter's position. These FFIs provide an opportunity to observe a given star in comparison to a set of nearby standard stars, and that is what \citet{Mont17} have done (see that paper for a full description of their methodology). This supplies 45 absolute flux points over the 4 years of observation, and these can be compared with knowledge of the level of the star's variability at those times (measured directly from the long-cadence pipeline light curves). 

\begin{figure}
    \begin{center}
    \epsscale{1.0}
    \plotone{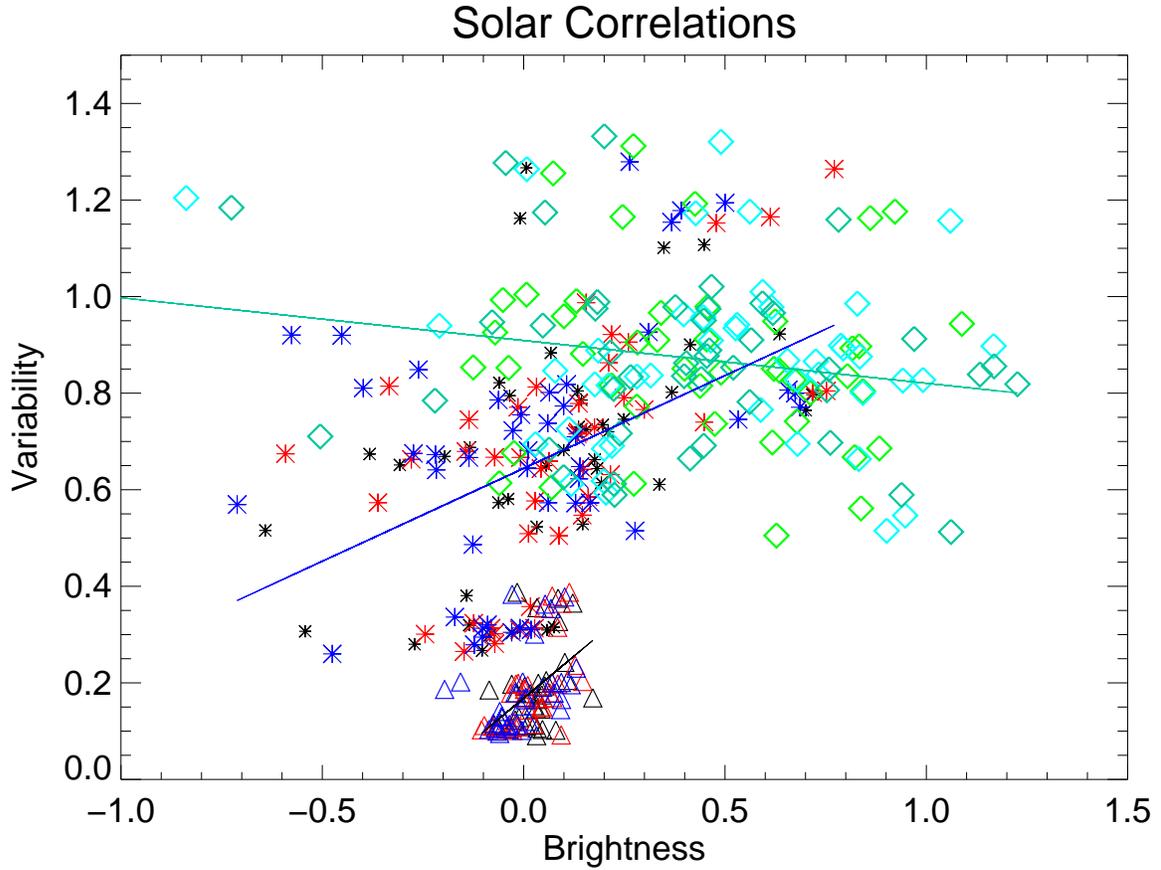}
    \caption{A comparison of absolute brightness with the level of variability at sets of times (see text) for the Sun. Both axes are in ppt; the absolute brightness is relative to the Cycle 23 median. Different colors for the same symbol represent different phasing of the measurement cadence. Diamonds are for the active Sun, asterisks are for the transitional Sun, and triangles are for the quiet Sun. There is a clear positive correlation between brightness and variability for the transitional (blue line) and quiet (black line) Sun, and a weak negative correlation for the active (green line) Sun.
    \label{fig:SunCorrl}}
    \end{center}
\end{figure}

For the Sun, this process can be mimicked by sampling the absolute solar curve at a similar (monthly) cadence, and evaluating the solar variability at those times from the full time series (as with Kepler). Sometimes those times will occur while the Sun is in a shallow or deep dip and sometimes when it is relatively bright. Of course, the phasing of this sampling is arbitrary, so we take 3 samples offset by 10 days each to get a fairer representation of what Kepler might have seen. Variability in this context refers to timescales of several rotation periods rather than long-term activity cycles. We measure it with the range in differential amplitude during each 90 days. Various measures of variability are discussed in the next section; all the methods discussed give similar results. The result of this analysis is shown in Fig. \ref{fig:SunCorrl}. There is a clear (positive) correlation between variability and brightness for the transitional Sun (asterisks), with a slope of 0.38 and a Pearson-r coefficient of 0.19.

Unfortunately the terminology about this relationship is very confusing, since this case is often referred to as ``anti-correlated". Perhaps that is because the result is counter-intuitive in the sense that one would naively think that greater coverage of darker features should lead to lower brightness. One can see in Fig. \ref{fig:TranSun}, however, that the absolute Sun is brighter when the spot features are larger and more numerous. That is the result of the ``invisible" bright faculae, which more than compensate for the spots. In fact, the Sun really is darker when large spots are facing us (at the bottom of the deep dips), but the integrated flux over time is brighter than when it is quiet. 

The same correlation between brightness and variability is present for the quiet Sun (triangles), with a larger slope (0.6) and r-coefficient (0.5) but a much smaller amplitude. This is less surprising, since mostly one is seeing the effect of more or fewer faculae; spots are almost absent (Fig. \ref{fig:QuietSun}). In that case, having more faculae should enhance both the brightness and the variability compared with the non-magnetic case. On the other hand, things are rather different for the active Sun (diamonds). The variability is of course generally higher than for the transitional Sun, while the absolute brightness also is generally higher and has a greater range. The relationship between brightness and variability is weakly in the other direction, with a slope of -0.04 and r-coefficient of -0.2. This means that the active Sun actually is weakly ``correlated"; greater spottedness goes with a darker flux (the intuitive expectation). None of these correlations is strongly statistically significant, but they are astrophysically interesting.

\citet{Shap14} provide the underlying explanation for these correlations, namely that as magnetic activity increases, the spot area grows faster than the facular area and spots begin to dominate the photometric signal at some point. Their Fig. 7 shows that the Sun is in a regime where it can switch behaviors and become spot-dominated at times of strong activity, and the analysis here explicitly supports this. The implication is that most of the Kepler stars with known rotation periods, which are for the most part at least as active as the Sun or more so, should be spot-dominated (``correlated"), with greater photometric variability accompanied by lower brightness. That is the result found by MTF17. This is a stronger argument against blanket application of the Flat methodology for deriving deficit curves. That method prevents the spot coverage from being related to the variability, because the coverage is inferred to be relatively constant regardless of variability. 

\subsection{Comparison with Starspot Models\label{sec:Models}}

The Sun does not have a light curve that resembles most of the Kepler stars with known rotation periods; they tend to be more variable than even the active Sun (which makes determining their rotation periods easier). In particular, the Sun never looks like the star in Fig. \ref{fig:Prob}, with very different levels and discrete envelopes of great variability. For that reason, it is important to test the methodologies with spot models that look more like many of the Kepler light curves. A future major paper will examine tens of thousands of models spanning a large set of model parameters. Here we utilize some of that work to present a (very small) representative sample of Kepler stars and models that resemble them. The basic modeling code is taken from \citet{Walk13}; the underlying scheme of analytic spot models is due to \citet{Dor87}. 

The modeling procedure has many relevant parameters that must be set, but the most important are the number of spots, their spatial distribution, how much (if any) differential rotation is present, and what lifetime (if evolution is allowed) the spots have. Evolving spots grow linearly for half their lifetime to their specified size, then decay linearly (not a physically accurate representation, but that detail is not important for now). At least 1000 trials are usually run for a given parameter set. For the examples presented here, spot location is fully randomized in longitude and visible latitude, the stellar inclination is 60 degrees (both the ``most likely" value and one which both allows for spots that go behind the visible hemisphere and spots that don't), the spot sizes are all the same (either 3 or 4 degrees in radius; they could be scaled to produce the depths present in an observed light curve), the rotation period (before differential rotation) is 10 days (this also could be scaled to an observed period), and the number of rotation periods computed is 40 or 80. 

We surveyed the residual differences in light deficit between actual model output and after applying the MidFit method computed with various time bin sizes for a large number of models. The best results (lowest standard deviation residuals) tended to occur with smoothing bins between 8 and 10 rotation periods and occasionally 6. Based on the full set of trials, our preferred temporal smoothing is bins of 8 rotation periods to determine the points of a spline fit to represent the overall variations of the continuum. Optimally the bin size should probably depend on the amplitude and frequency of large variations in the differential light curve. These trials also made it clear that the CoarseFit and FineFit methods are worse choices in almost all cases.

Out of the many sets of trials we selected parameters that generated light curves qualitatively similar (in the frequency and relative depth patterns of greater and lesser variability, and to a lesser extent in the occurrence of single/double dips) to the Kepler data. The spot lifetimes and extent of differential rotation, along with the number of spots present, are the controlled variables. The examples shown below were all chosen from the first 50 trials in a given set of 1000, which means they are fairly typical of the set. It should be pointed out, however, that each parameter set also contains light curves quite different from those shown here. The random distribution of spots in longitude makes a big difference. One advantage of the models is full information on the location and size of every spot all the time (visible or not) and the deficits they each cause.

\begin{figure}
    \begin{center}
    \epsscale{1.0}
    \plotone{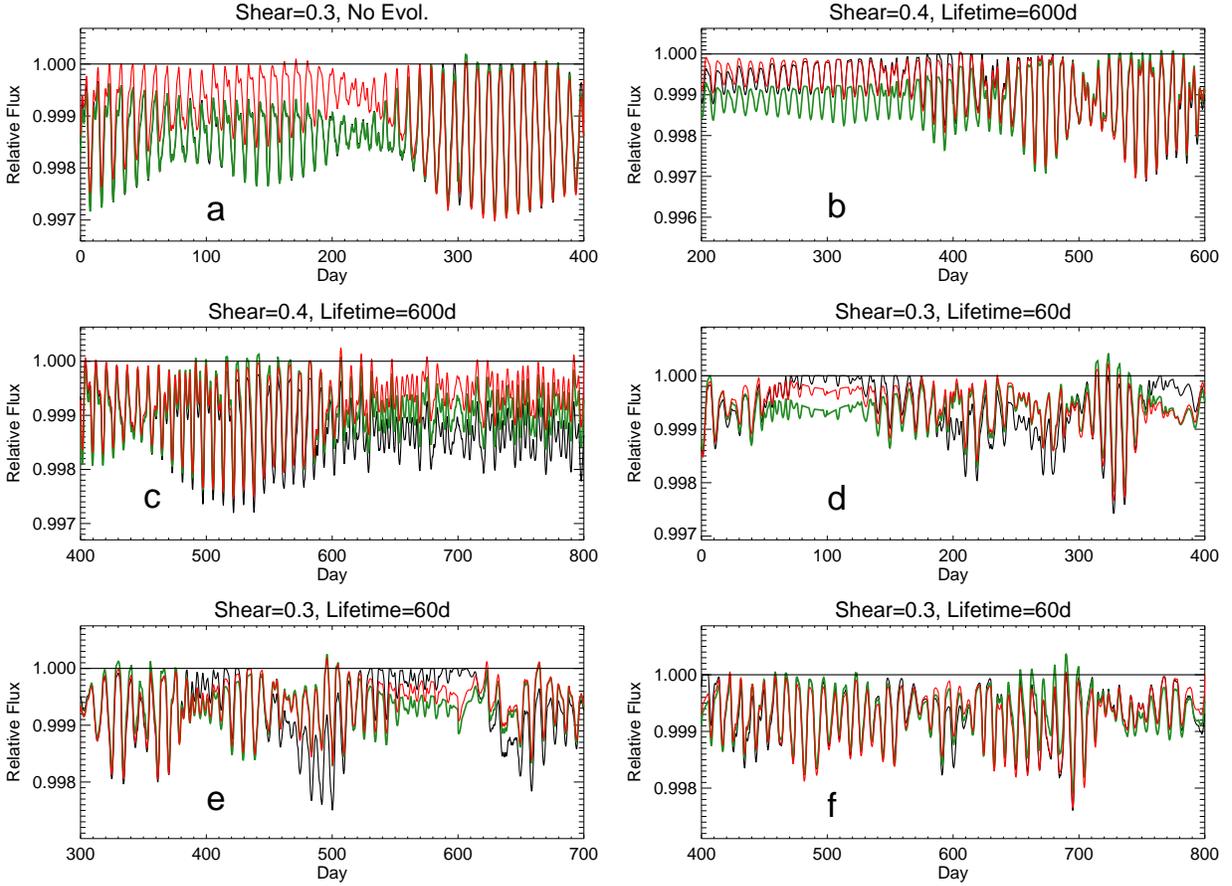}
    \caption{Examples of starspot models and their calibration with the Flat and MidFit methods after conversion to Kepler-like form. Panel (a) has pure differential rotation, (b,c) come from shared model parameters dominated by differential rotation; (d,e,f) come from shared model parameters dominated by spot evolution. Each instance has a different random set of spot positions and birth dates. The black curves are the original model output, red curves are the MidFit solution and green are the Flat solution, each applied after ``Keplerization".
    \label{fig:ModExamp}}
    \end{center}
\end{figure}

Working from models means that the true level of spottedness (coverage) is known, and the model light curves can be compared before and after they have been treated to resemble Kepler pipeline reductions (``Keplerization"). The procedure for that treatment is very similar to what was used for the Sun above, with one difference. A simple renormalization to the quarterly median is enough for the Sun because the coverage does not change greatly on that timescale. The models all have periods of 10 days (rather than 27), so their coverage can change more drastically during 100 days (the window used to define ``quarters"). We therefore fit third order polynomials to the quarterly light curves and divided by them to ensure that the median of the differential light curve is nearly constant (as it is in Kepler data). This resembles what was done by \citet{Bas11} before much was understood about the Kepler pipeline products. In each of the panels of Fig. \ref{fig:ModExamp}, the black curve shows the model output (truth), the red curve shows the application of the MidFit method after Keplerization, and the green shows what happens if instead the Flat method is applied after Keplerization.

The discussion above pointed out that using the Flat method is more appropriate if spots do not evolve, since that means the coverage is constant which is an strongly favored implication when using the Flat method. If the light curve does change from rotation to rotation given constant coverage, that must be due to changing spot spatial distributions. Without evolution, the only way to alter the distribution of spots is to move them around (differential rotation). The model in panel (a) of Fig. \ref{fig:ModExamp} illustrates these characteristics, being a case of pure differential rotation (with 50\% more than the solar shear) using 5 spots with no evolution Needless to say this is an idealization; real stars can't support permanently unchanging spots. 

The Flat solution (green) almost perfectly reproduces the original model (black), which means that Keplerization does not do violence to it, and gives hope that the pipeline doesn't either (at least for cases with many periods per quarter). The lower amplitude segment (before day 250) is well under the continuum, reflecting that there is plenty of spot coverage but the spots are more evenly spread around the star. This is because the continuum is set by the brightest points in the Keplerized version, which occur where the differential amplitude is greatest (after day 300). In fact, the spot distribution is sufficiently asymmetric then that the model just reaches unity at these points; the star is briefly unspotted. The methods several authors have used to identify stellar activity cycles would be fooled by the light curve (especially the MidFit version) in panel (a), which looks cyclic but actually has unchanging spot coverage throughout (the spots are simply geometrically re-arranged). 

The MidFit calibration removes part of the coverage information in this case because it sets a more local continuum, making the coverage appear smaller in the less variable segments. Of course, if one didn't know that the light curve is purely due to differential rotation this would seem quite reasonable, and indeed in other cases like panel (b) of Fig. \ref{fig:ModExamp}, the MidFit method is more correct than the Flat method. The most egregious case of Flat method failure is the first half of panel (b), but there are also clear problems in panels (d) and (e). It is worth pointing out that unlike the Sun, model spots are placed at all latitudes, including those that are always visible (the Sun is viewed equatorially, so does not have any permanently visible spots). This allows the full range of differential rotation to be imprinted on the light curve. Given the restricted latitude range of spots at a given time on the Sun (and probably on similar stars) it is harder to detect differential rotation because it is differential across latitude. In addition to the copious evidence for polar spots on rapid rotators \citep{Stras09}, there is direct evidence for numerous high latitude spots seen by transits \citep{Morris17}.

The failure of the Flat method in the first half of panel (b) is caused by the fact that although this model has even more differential rotation than panel (a), it also allows for a little spot evolution. The spot lifetime is very long (60 rotations) so differential rotation still dominates. This lifetime is nothing like the Sun's, but there are cases of young rapidly rotating stars that might exhibit such long lifetimes, for example \citet{Innis88}. In our evolutionary models, birthdays are assigned with uniformly random dates (including up to one period before the run starts). The model shown here happens to have less coverage before about day 350 than after due to its particular random choice of spot birthdays. The model has bright points near the continuum throughout because it is not too spotted, but the dips are smaller initially (when there are fewer spots) then grow and begin to show more of the ``beating" that differential rotation tends to produce as a few more spots appear (the total number of spots used throughout is 13). The Flat solution is influenced by the larger amplitudes near the end, while the MidFit solution is nearly correct throughout because of its somewhat local character. 

Panel (c) of Fig. \ref{fig:ModExamp} shows another case from the same parameter set as (b), but here the MidFit method fails after day 600 because the coverage in the model remains high even as the differential amplitudes are lower in the latter half. This is because the differential rotation has spread the spots more evenly around the star; for this model the coverage is more constant throughout than case (b) because of the more even random distribution of spots temporally. The Flat solution is intermediate between the model and the MidFit solution; this is due to the vagaries of the changing coverage. There can easily be cases where the model always has enough coverage that the peaks stay well below unity (if the whole run is like after day 600). This can easily be accomplished with some combination of numerous and large and/or polar spots. That is, spots can generate a persistent signal either because they are always in view due to inclination, or some base amount of coverage is always present because the star has enough spots everywhere. Then the Flat and MidFit methods may agree more with each other, but both give a significantly lower than true value for all the coverages. This is another instance in which one must acknowledge that derived coverages can be lower limits due to a persistent (and therefore unmeasured) deficit.  

The last 3 panels in Fig. \ref{fig:ModExamp} are for models in which the spot lifetime is much shorter (60 days). For these, spot evolution is dominant over differential rotation. A hint of a promising morphological characteristic of these light curves is evident: the ``beating" appearance is shorter, more erratic, or absent. Unfortunately that is not always the case among the 1000 trials of this parameter set, but it may be something to pursue. Panel (d) exposes another sort of problem. When the model has significant unspotted segments, for example at days 50-150, the Keplerization can run into trouble. The nature of the failure of the Flat solution depends in part on where the quarter boundaries fall; days 100-200 in panel (d) have a ``V" shape (green) because the model light curve (black) is rising in the first 100 days and falling in the second 100 days. The MidFit solution does a better job because of its more local nature, but can't fully correct for this. A less egregious example occurs in days 550-620 in panel (e). These cases are also a violation of the general dictum that the differential light curve provides a lower limit to the true coverage; in these examples the true coverage is even smaller than the inferred coverage.

Long flat sections do not occur in any of the observations with significant variability (not even as a steady slope to flat-topped peaks, as here). In a way that is a pity, since such segments would be excellent indicators of the true unspotted level (and thus could be properly calibrated). Of course, the transitional Sun does not show any flat continuum (Fig. \ref{fig:TranSun}); even the quiet Sun doesn't! It seems likely that when real stars are unspotted at length, they don't produce much detectable spottedness (unlike these noiseless models). Presumably the very large group of stars for which rotation periods cannot be measured by Kepler because they show no variability at all are basically unspotted. When there aren't long flat sections in the models, the MidFit solution often does a good job of coming close to the model, as in most of panels (b), (e), and (f), and (d) between days 150-350. Indeed, if the coverage also doesn't change too much, both methods are reasonably successful, as in panel (f). 

\subsection{Comparison between Observations and Models\label{sec:Obs}}

We now compare a few model solutions with actual Kepler observations. The four examples here are chosen by eye purely for pedagogical purposes. Each has a Kepler light curve and model that approximately match in qualitative appearance. The red curves in all the examples (for both observations and models) are the result of applying the MidFit method to either the pipeline light curve or the Keplerized version of the model to determine the continuum. The green curves for the observations are the result of using the Flat method instead to calibrate the Kepler differential light curves. The black curves in the model plots are the actual model output so they reflect the true unspotted level (unity) and the correct coverage at all times. The overall coverage levels of the model deficits were not matched to the observations, but could be rectified (with no gain in relevant information) by choosing a larger model spot size.

\begin{figure}
    \begin{center}
    \epsscale{1.0}
    \plotone{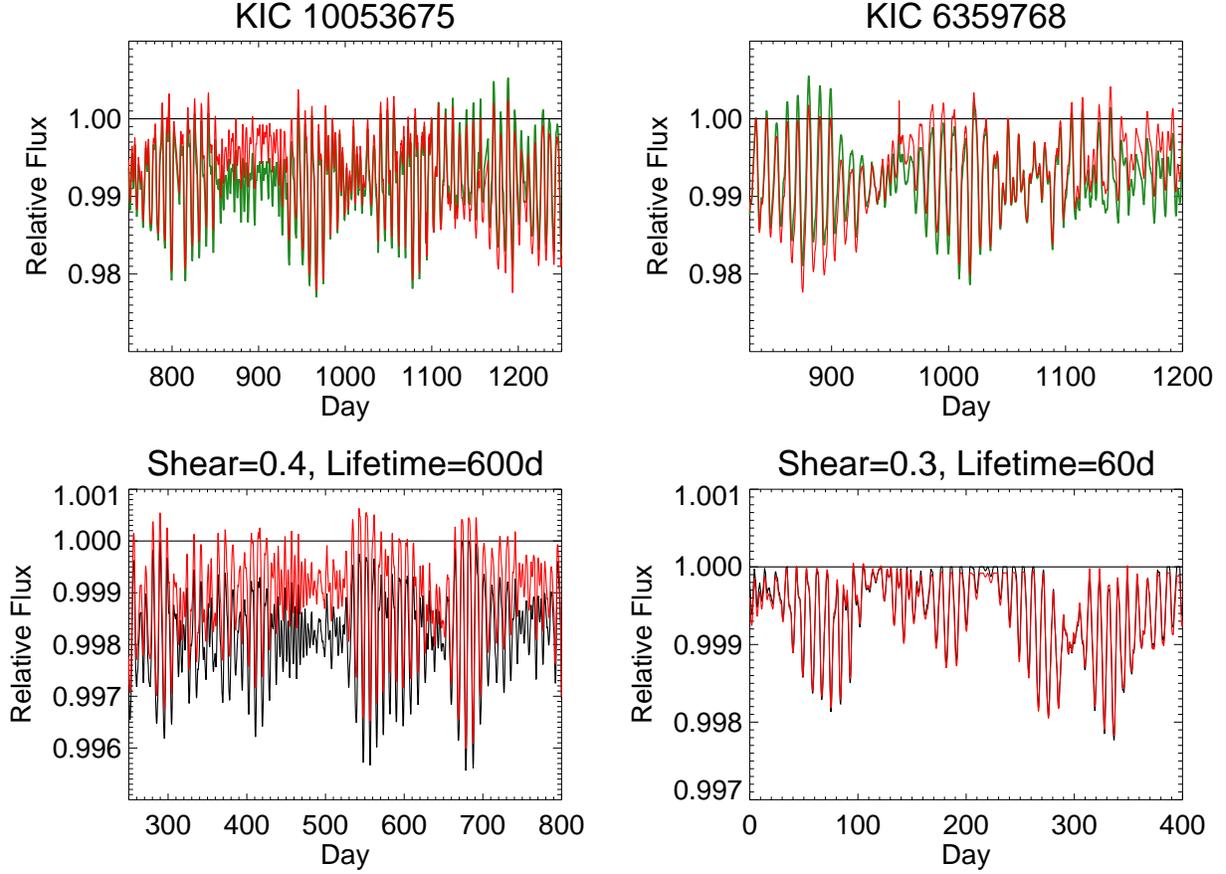}
    \caption{A comparison of two observations (upper panels) with two models (lower panels). The models were chosen to have qualitative similarities to the observations. The left model illustrates almost pure differential rotation, while the right model is dominated by spot evolution effects. The red curves in all panels are the result of using the MidFit calibration method (on observations or Keplerized model curves). The green curves in the observations are calibrated with the Flat method, while the black curves are the actual model output (before calibration or Keplerization). In the lower right panel, the MidFit solution is almost perfect.
    \label{fig:Examp1}}
    \end{center}
\end{figure}

The left panels of Figure \ref{fig:Examp1} show a case where there is a repeating oscillation between greater and lesser variability. On top is KIC 10053675, which has a rotation period of 8 days. The model below it has the same parameters as for panels (b,c) of Fig. \ref{fig:ModExamp}. The model is unlike the Sun in several respects: spots can appear at any latitude (which enhances the observable shear), there is more than twice as much coverage as on the active Sun (but much less than the observed star), sunspots only last less for 1-2 rather than 60 rotations, and it is viewed at a 60 degree inclination. It has about 10 spot groups present at any given time. Presumably the solar-type star being observed also has spots quite unlike the Sun, since it looks like the model. This is due to its rapid rotation.

\begin{figure}
    \begin{center}
    \epsscale{1.0}
    \plotone{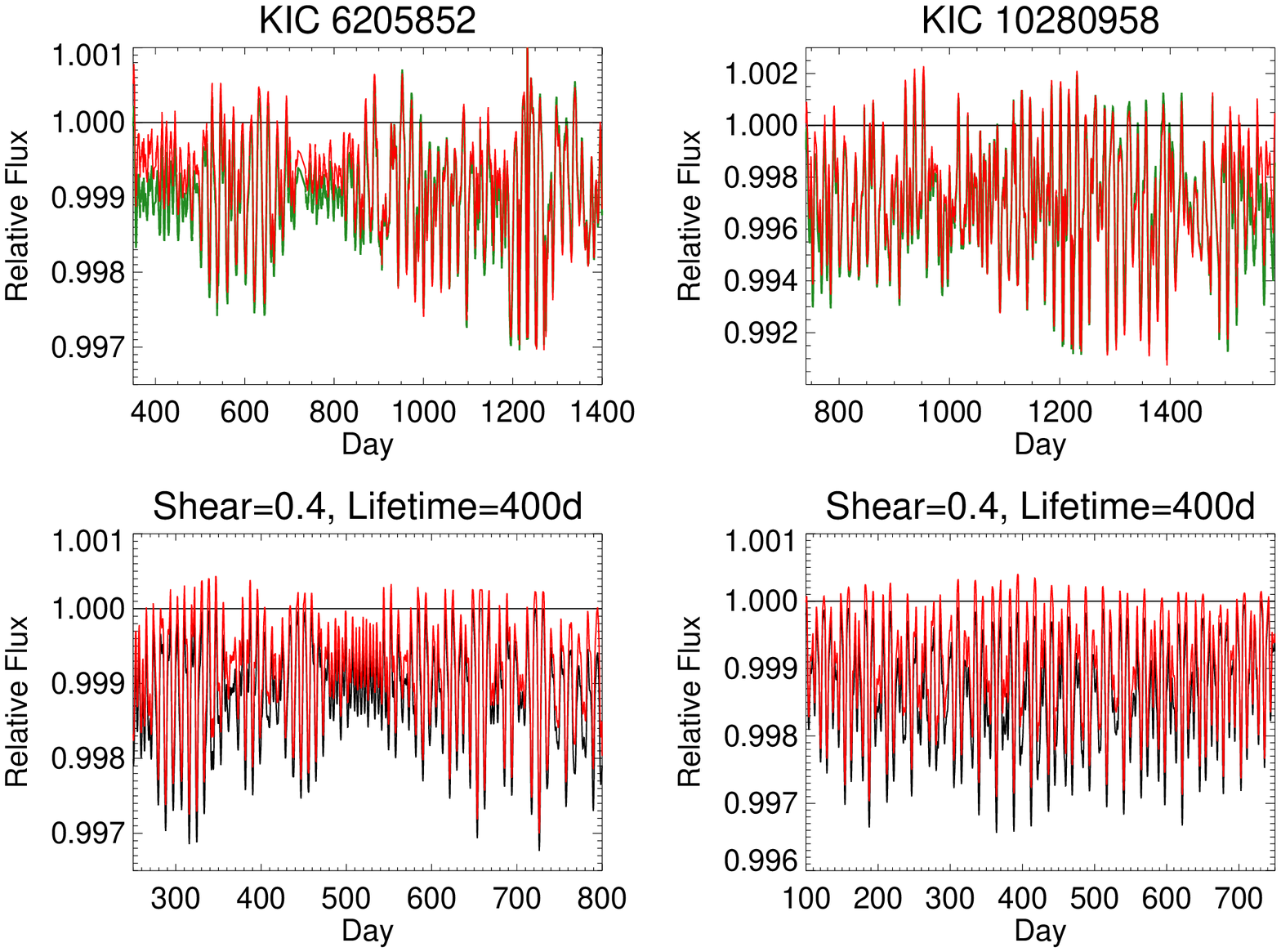}
    \caption{Another set of observations and models, similar to Fig. \ref{fig:Examp1}. For the right observation, the Flat and MidFit solutions are quite similar. Both model have the same parameters (strong differential rotation and slow spot evolution); only the random spot distribution is different. The right set shows a case in which there are mostly single dips, and no coherent patches of greater variability.
    \label{fig:Examp2}}
    \end{center}
\end{figure}        

The discussion above of panel (a) in Fig. \ref{fig:ModExamp} also applies to the left panels of Fig. \ref{fig:Examp1}. A mismatch between the Flat and MidFit solutions arises in the observation between days 850-950, but of course one doesn't know which version of the observed light curve is more accurate (it should be the green curve if the star is exhibiting mostly differential rotation). A similar mismatch occurs in the model between days 320-550. Although hard to discern in the plot, some of the large amplitude dips occupy a whole rotation period, while there are 2 dips per rotation in the lower amplitude segments and also in part of the higher amplitude segments (the sections that appear ``denser"). This is an example of the single-dip/double-dip dichotomy discussed by \citet{Bas18}. They defined the single/double ratio (SDR) as the logarithm of the total duration spent single over the total duration spent double, so it is positive for predominantly single light curves and negative for predominantly double ones. Similar features are seen in the model, the SDR is -0.18 for the observation and approximately zero in the model (which was not chosen to match this particular characteristic). Notice that the lower variability segments have double dips, which is expected from differential rotation \citep{Bas18}. They also note the possibility this could arise from evolution; if the star suddenly had fewer spots during these times that could produce this effect. Since the model has slow evolution, however, the cause is definitely differential rotation there.

The right panels of Figure \ref{fig:Examp1} show a case in which the light curves still show some ``beating" (alternating segments of greater and lesser variability), but the model has a much shorter spot lifetime (and a little less differential rotation). Unlike the previous case, the amplitude fluctuations in the model are now primarily due to changing spot numbers, and the coverage changes accordingly. Here the observation has an SDR of 0.06 while the model has 0.44 (nearly 3 times as many single dips). In both instances the difference between the Flat and MidFit calibrations is significantly less because high variability segments occur throughout the light curve. There is still a segment near the end of the observation where the MidFit version has a little less coverage than the Flat version and the opposite occurs near the beginning. 

It is not possible to confidently claim that KIC 6359768 shows primarily evolutionary rather than differential rotation behavior, although it is tempting to do so given its similarity to the model dominated by evolution. There may be some promising ideas for making such a determination related to the morphological characteristics pointed out in this discussion (the nature of the beating; single/double behavior), but a lot of work remains before one could be sure of such a conclusion. Until these issues are better understood, assigning bulk behavior of the variability changes to details of differential rotation (as for example \citet{Rein15} do) must be considered somewhat uncertain.

Two other cases are presented in Figure \ref{fig:Examp2}. These models have large shear and long (40 rotations) spot lifetimes, but both models come from the same parameter set. This illustrates that the actual (random, different) distribution of spots can really matter. Each model has about 5 spots present at any given time; they use 10 total (including replacements for spots that died during the 80 rotations). The left panels show observations and a model run that have large changes in variability; the SDR for KIC 6205852 is about 0.3, while for the model it is -0.2. On the right there is more constant high variability. The SDR is 0.57 for KIC 10280958 and 0.415 for the model, so they are both primarily single-dip cases. The only difference between the left and right models is the random distribution of spots; there are the same amounts of both spot evolution and differential rotation for these two cases. Nonetheless there are more coherent light curve structures in the left case, while the right case exhibits more consistent levels of variation with little longer term structure. 

It should be clear by now that {\it there is no one surefire method for deriving the true coverage from differential light curves}, particularly if one cannot separate out effects of differential rotation from effects of spot evolution sufficiently well. Results like those in Figs. \ref{fig:Examp1} and \ref{fig:Examp2} serve as a warning that accurately inferring physically relevant spot behaviors from differential light curves is not an easy task, even for global properties like the amount of differential rotational shear or spot lifetimes (not to mention the properties of individual spot groups). This is even before consideration of the uncertainties about what the Kepler pipeline does to brightness trends (of various sorts) on intermediate timescales, although we will argue below that is not the main problem. 

On the other hand, in many cases (when they produce the same result) it doesn't matter too much which calibration scheme is employed to produce a deficit curve from a differential light curve. This most true if the variability doesn't change too much, or the segments of low and high variability are sufficiently interspersed. There is a general tendency for the Flat method to yield greater inferred coverages, especially when the variability is lower for a long stretch. When differential rotation is dominant these greater coverages are actually more correct. Unfortunately, at the moment it is not generally possible from the observations to determine which set of coverages is more appropriate. 

Based on these cases and theoretical modeling of many more spot configurations, and recalling Fig. \ref{fig:Prob}, it is clear that a low order fit to the unspotted continuum over the whole light curve is generally too coarse and a fit following fewer than about 4-6 rotational peaks at a time is too fine. The greatest uncertainty occurs when the variability goes through large changes, especially over long stretches of time. It is usually better to use the MidFit method to convert a differential light curve into a light deficit curve unless one is convinced that differential rotation dominates, in which case it is best to use the Flat method. At the moment, however, there seems no way to insure that significant errors can be avoided in all segments. What is most helpful is to apply both methods. When they agree with each other, it is much more likely safe to proceed with the derived deficit curve.

\section{Kepler Deficit Curves, Absolute Photometry, and Faculae \label{sec:analysis}}

Armed with a methodology for converting differential curves into deficit curves, we turn to the question of whether faculae also need to be considered in properly calibrating Kepler light curves for starspot analysis. We employ the stellar sample utilized by \citet{Mont17} (MTF17); a set of stars that show coherent variability in the absolute photometry measured by those authors. This is comprised of 463 solar-type stars, within 150K of solar temperature and with gravities of $log(g)>4.2$ (from the catalogs at the time MTF17 picked them). They range in Kepler magnitude from 10 to 16. The data used in the current analysis are PDC-SAP products from the last release of the Kepler project pipeline (DR25). The long cadence Kepler data have a 29 minute time resolution, which is far finer than needed to track the variability of spots. The data were therefore rebinned by a factor of 10 to timesteps of about 0.2 days, improving the S/N and eliminating rapid features unlikely to be due to starspots. The pipeline produces a record of differential intensities that (after dividing by the quarterly medians) oscillate below and above unity with an amplitude that can be anything from about $10^{-4}-10^{-2}$ (representing photometric variations of a few tenths of a millimag to a tenth of a magnitude).

\begin{figure}
    \begin{center}
    \epsscale{1.0}
    \plotone{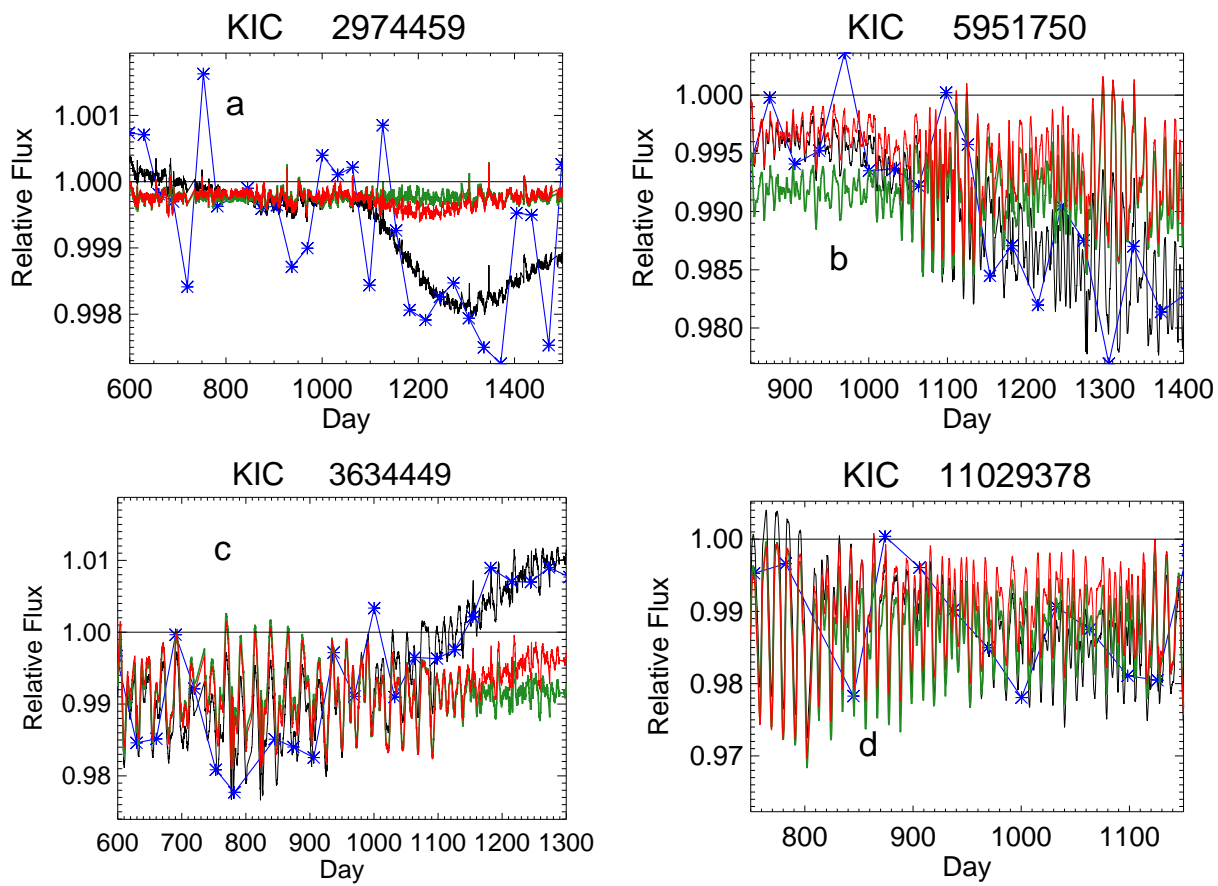}
    \caption{Calibration with Absolute Flux. Each panel shows an observation including the absolute flux measurements. The black curve shows a calibration of the pipeline differential curve produced by forcing it to follow a smoothed fit to the absolute flux measurements (shown in blue). The red curve shows the result of converting the black curve to a deficit curve using the MidFit method; the green curve uses the Flat method. 
    \label{fig:CalExp}}
    \end{center}
\end{figure}        

The first question is whether applying the absolute fluxes from MTF17 helps with calibrating the differential light curves. Those authors realized that because the FFI frames were taken at discrete times, the pipeline light curves provide a snapshot of what the star was doing just before and after the FFI was taken. Thus, if one knows that the star was in the middle of a deep dip, it is not surprising if the absolute flux taken then is lower than it might otherwise be. They corrected their fluxes so that the measurements given refer back to the average flux for the quarter. They also did their analysis with uncorrected fluxes, and found the same final results. Because we are interested in correcting the light curves onto an absolute flux scale (not preserving the quarterly mean), their uncorrected fluxes were taken as the starting point. 

It was readily obvious that one cannot simply force the light curve to match onto each of the FFI points, since they regularly display changes with an amplitude that is not supported by the behavior of the long cadence differential light curves. Panel (a) of Fig. \ref{fig:CalExp} shows an extreme example (only a few stars have such striking behavior). This is a bright star (Kep$_{mag}=12.5$). The differential light curve is very quiet throughout, but the absolute fluxes are much noisier and show a puzzling long term dip in the later portion. Section \ref{sec:bvcorrl} contains more detail about noise; the total absolute photometric scatter (both instrumental and stellar) for this star given by MTF17 is 1.2 ppt. The rms noise in the pipeline long cadence light curve is only 0.03 ppt (with a full range of 0.3 ppt). 

Forcing the differential light curve to follow the absolute (blue) points in detail is incompatible with the quietness of the differential observations. In order to preserve some of the information on intermediate timescales a triangular smoothing of 5 points was used on the fluxes from MTF17, which occur at about one month intervals. The resulting smooth functions that characterize the trends in the absolute points are obvious from the black curves in Fig. \ref{fig:CalExp} (which follow them). It is beyond the scope of this paper to speculate whether the long-term trend absolute trend in panel (a) is itself problematic. If one wished to follow only the very longest term trends one could fit a low order (third order, for example) polynomial to the absolute points. That would convert the dip at around Day 1300 in panel (a) to simply part of a general decreasing trend. 

Panel (b) shows a portion of the same light curve shown in Fig. \ref{fig:Prob}. It is clear that the first part of the curve is less variable and brighter with a smaller deficit, while the last part is more variable and fainter with a larger deficit. The same is basically true in panel (c), and these examples are more typical of the larger sample of stars. Panel (d) illustrates one of the relatively few stars with the (mildly) opposite behavior: the more variable first part is brighter and has a larger deficit, while the last part is fainter but less variable and with less deficit. 

It would make no sense to use the Flat method after applying the absolute calibration; that would presume that the large brightness excursions are entirely due to changes in spot coverage, and ignore the facular information that is presumably gained by the using absolute calibration. It also seems that the MidFit solutions (red curves) are generally more plausible than the Flat method produces in most cases, if we take the absolute variation trends seriously. 

The Flat solution applied to the differential light curve tends to yield a relatively constant spot coverage over time, as demonstrated in Section \ref{sec:Models}, regardless of what the absolute fluxes are doing. In particular, panels (b) and (c) would have the same spot coverage in the less variable segments and more variable segments using the Flat method, but the less variable ones are brighter in absolute flux. A physical scenario that could produce that result is not easily apparent. It would require that faculae somehow become dominant if spots are spread more evenly around the star. The interpretation from the MidFit solution is more straightforward: the star gets darker as the spot coverage becomes stronger.

Not apparent in Fig. \ref{fig:CalExp} is the fact that it makes very little difference whether the MidFit conversion to a deficit curve is done on the light curve before or after applying the absolute calibration. The changes in slope due to the absolute calibration are gentle enough (admittedly because of the smoothing needed and used) that the process of fitting a continuum with the MidFit method yields essentially the same result with or with them. This is good news, since most Kepler light curves do not (yet) have absolute calibrations. It means that {\it one can infer the deficit from pipeline light curves with our methods to their full efficacy without requiring the FFI photometry}. The bad news is that the facular information is therefore not easily recovered from the light curves, or put more positively, the facular signal is relatively independent of the spot signal and is not required for the spot analysis. That is essentially also the conclusion reached for the Sun in Section \ref{sec:Sun}.

To summarize, the best treatment of pipeline differential light curves to prepare them for starspot analysis at this time appears to be to use both the Flat and MidFit methods to find the unspotted continuum reference from which the light deficit curve is measured (and hope they agree). This can be done without knowledge of the absolute photometric variations due to faculae. When they don't agree, an assessment must be made on which to use. If one believes (for reasons that require further development) that the light curve is dominated by differential rotation effects, it is better to use the Flat method, otherwise the MidFit method is safer. This procedure yields lower limits to the true starspot deficits, since it is unaware of any persistent starspot signal present in addition to the differential changes. 

\subsection{Brightness-Variability Correlations\label{sec:bvcorrl}}

\begin{figure}
    \begin{center}
    \epsscale{0.75}
    \plotone{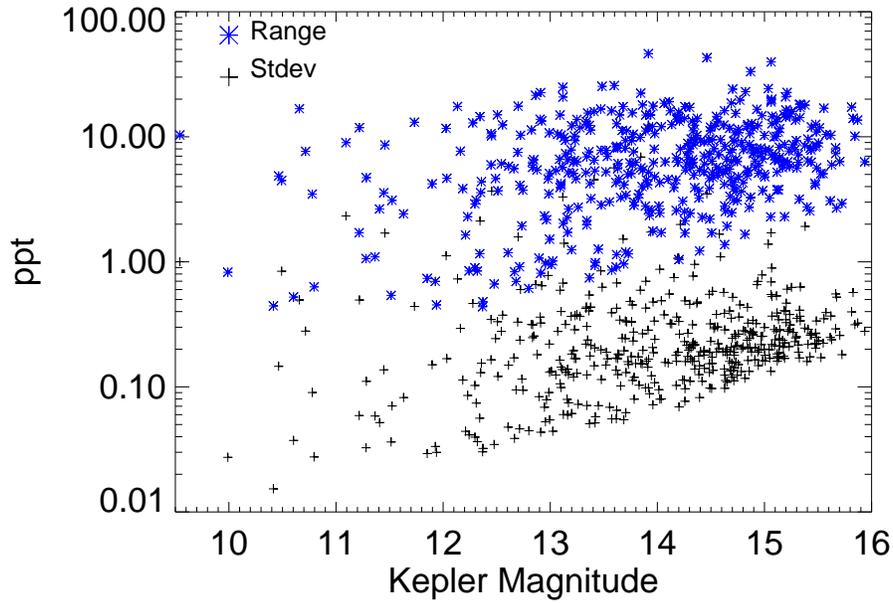}
    \caption{Instrumental and Stellar Variability. The plusses show the local residual variability (defined in the text) of stars as a function of their apparent Kepler magnitude. A floor is apparent which rises to about 0.2 ppt at $K_{mag}\approx$ 16. Most targets show a larger value that arises from stellar variability. The asterisks show the median quarterly differential photometric range, which is typically more than an order of magnitude larger due to longer timescale variability. 
    \label{fig:Noise}}
    \end{center}
\end{figure}

The correlations between brightness and variability in the calibrated light curves were explored by MTF17, who drew some broad conclusions about the relative importance of spot and facular contributions. It has been established using ground-based photometry and spectroscopic activity measures that there are two classes of solar-type stars: the faculae- and spot-dominated cases \citep{Hall09}. MTF17 utilized a fairly local measure of variability then correlated it with the absolute flux measurements. We look at this question in more detail to study whether there is a facular signal that is relevant to the deduction of correct spot coverages. In order to evaluate the S/N for the variability analysis, we compute a noise metric by smoothing the binned data with a boxcar of 4 timesteps, then subtracting the smoothed curve from the original. The standard deviation of this residual is the noise metric, although it is not strictly speaking a true noise term. That is because some stars have intrinsic variability on timescales approaching the smoothing \citep{Bast13}, so some stellar variability is included in the metric in those cases. As can be seen in Fig. \ref{fig:Noise} there is a floor on this metric which rises with Kepler magnitude; that is a better indication of the extent to which real noise is contributing to this metric. 

The metric for stellar variability adopted here is the ``range" ($R_{var}$) as defined by \citet{Bas11}. This is the difference in a differential light curve between its 5\% and 95\% brightness levels (after sorting all points by relative brightness). It can be computed on various timescales; we first consider the median of the quarterly ranges for a given light curve (as used by several authors). These median ranges for the sample are also shown in Fig. \ref{fig:Noise}; they generally lie more than an order of magnitude above the noise metric. They show a similarly rising lower envelope, so some of the stars have low enough $R_{var}$ that they are also clearly affected by photon noise. There are also some exceptional points where the ``noise" and $R_{var}$ are comparable (though $R_{var}$ is still at least a factor of 3 greater for a given star). These stars have very rapid periodic variations; all but two of them have inferred periods under 3 days (and the variations may be pulsations rather than rotation). They were checked for special behaviors that might affect the analysis or conclusions below; none were found.

Other measures of variability were also considered. One is to cut the light curve into segments centered on each of the absolute calibration time points from MTF17 and extend them halfway to the next point on either side, then measure the range for each of these segments. This doesn't require knowledge of the rotation periods. The ``Golden FFI" (MTF17) points taken during commissioning were lumped together, and because they occur before the beginning of the long cadence light curves, their average value was interpolated along with the next absolute flux value onto the beginning epoch of the long cadence curves. A more local measure of variability is the range over one rotation, $R_{loc}$. It can be measured along with integrated coverage once the timesteps for each rotation have been identified. Obviously this measure can change rapidly so we first smooth the full set of $R_{loc}$ vs time values over 6 rotation periods, then linearly re-sample that onto the absolute calibration epochs. A third metric is that adopted by MTF17 (called $S_{ph}$). It is defined as the standard deviation of the differential light curve over each 5 rotations \citep{Math14}. In the analyses below, all three of these measures were tried and found to be essentially equivalent for the purpose of correlating brightness with variability. The method actually employed in this paper is the re-sampled fit to the smoothed $R_{loc}$; we will call this $R_{smo}$. 

\begin{figure}
    \begin{center}
    \epsscale{0.65}
    \plotone{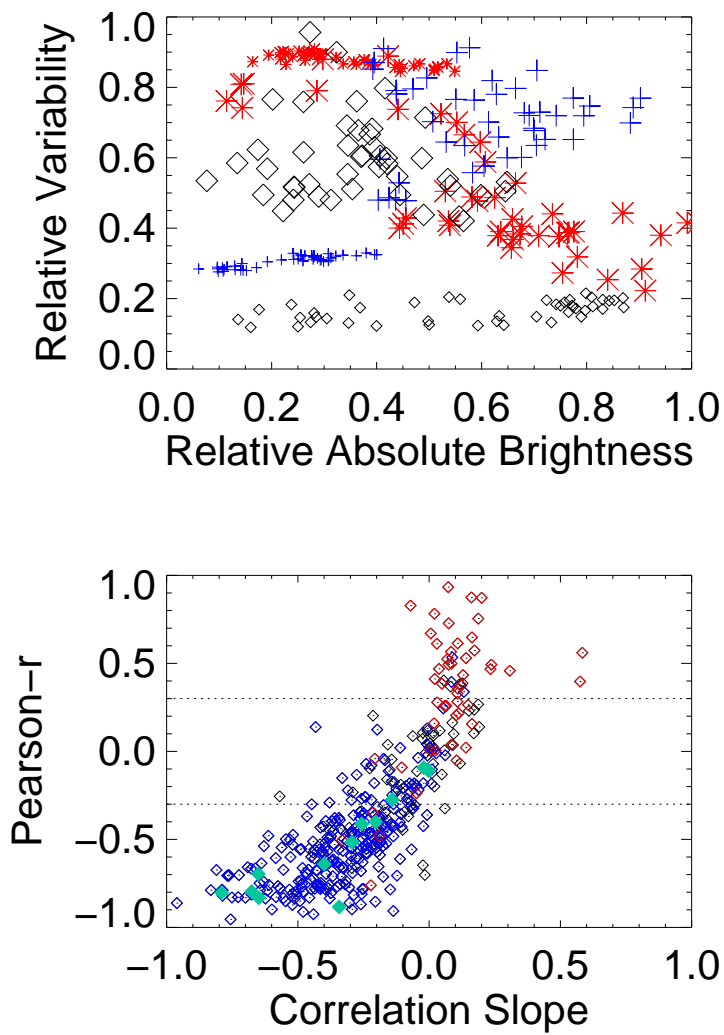}
    \caption{Upper Panel: Five illustrative sets of brightness versus variability. They have been scaled and shifted to fit on the plot. Color and symbol size are related to slope (see text).
    Lower Panel: Slopes of the linear fits to the brightness versus variability points for all stars; the ordinate is the Pearson-r coefficient for each star. The colors indicate the designations from MTF17 based on the sign of their correlation slopes (see text). Green points are stars with clear brightness cycles in MTF17.
    \label{fig:CorrlComp}}
    \end{center}
\end{figure}

This leads to a set of both absolute brightness and variability measurements at the same 45 time points in each light curve. A few illustrations of these sets are provided in the upper panel of Fig. \ref{fig:CorrlComp}. A linear fit is made to each set of points and both the slope and Pearson-r correlation coefficient are computed. For the small and large red asterisks the [slope, Pearson-r] values are [-0.14,-0.91] and [-0.92,-0.84], for the small and large plusses they are [0.16,0.87] and [0.59,0.53], for the small and large black diamonds they are [0.00,0.01] and [-0.20,-0.30].  The KIC numbers for the stars shown (in the same order) are:  5302725, 9334867,  8078753, 11014223, 2694810, and 11299398. Shown in aqua in the lower panel are the 12 stars which MTF17 identified as exhibiting clear activity cycles over the 4 years of observations (their Fig. 5). Four of these are very clear-cut cases of spot-dominance  (those with the most negative r-coefficients and slopes), two are unclear (those above the lower dotted line), and the rest are distributed along a continuum of spot-dominance.

MTF17 divided the sample into 3 cases: anti-correlated (red), correlated (blue), and not clear (black); the colors in Fig. \ref{fig:CorrlComp} correspond to their classification. As explained in Section \ref{sec:Sun}, the ``anti-correlated" label is used for stars which become brighter as they become more variable; they are generally interpreted to be faculae-dominated. The points for such stars have a positive slope, so they should lie in the upper right quadrant of the lower panel of Fig. \ref{fig:CorrlComp}, which shows the results for the full sample. Similarly, the blue (correlated) stars get darker as they get more variable (spot-dominated), and are in the lower left quadrant. Most stars from the sample (320 in MTF17; 322 for us) are in this category. In most cases (310 stars) their sense agrees with ours although there are a few red points in the lower left quadrant, and stars which they didn't classify in either category (black) are found in both quadrants for us. The most likely source of discrepancies is that here the individual absolute fluxes are smoothed over several months unlike in MTF17.  

In social science a correlation coefficient with absolute value of less than 0.5 is not considered convincingly significant. In examining the individual sets of points for stars in that regime, many of them looked (subjectively) convincing enough for this purpose, so we classify stars as unclear or uncorrelated if their Pearson-r coefficient has an absolute value less than 0.3 (between the dotted lines in the lower panel). The large black diamonds in the upper panel of Fig. \ref{fig:CorrlComp} illustrate the quality of a correlation barely below the cutoff. Of course, points that exhibit little slope will have small r-coefficients even if they have low scatter (like the small black diamonds). We also computed two other statistical tests: the Spearman $\rho$ and Kendall $\tau$ coefficients, and reached very similar conclusions. The threshold of significance adopted is rather generous; assigning non-significance to a larger portion of the sample could easily be justified. Many of the stars in the upper right quadrant don't have a pronounced positive slope ($<0.2$) even when their correlation coefficient is respectable. 

\begin{figure}
    \begin{center}
    \epsscale{0.65}
    \plotone{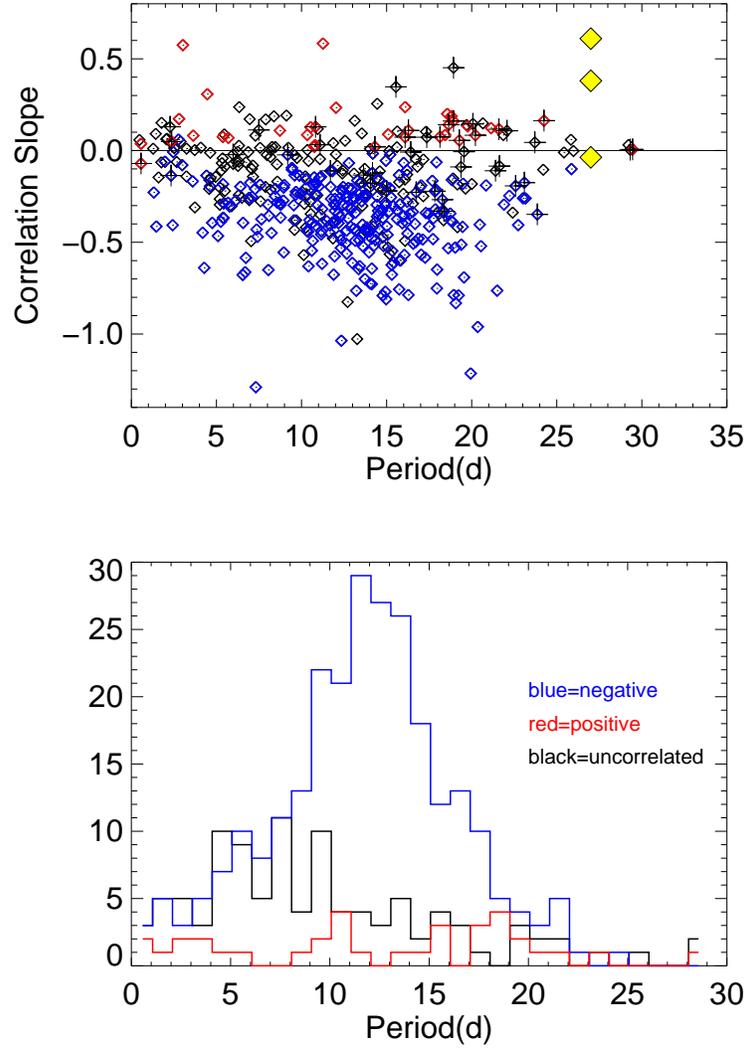}
    \caption{The upper panel shows the brightness-variability correlation slopes versus rotation period. The blue points have a Pearson-r coefficient of less than -0.3 and the red points have it greater than 0.3 (the black points lie in between, and so are not convincingly correlated). Crosses indicate points near the the noise floor in range. The correlations for the active, transitional, and quiet Sun are shown in yellow. The lower panel shows histograms of the distributions of the 3 classes of correlations with rotation period.
    \label{fig:PerCorrl}}
    \end{center}
\end{figure}

The upper panel of Fig. \ref{fig:PerCorrl} shows the correlation slopes as a function of rotation period. Points with crosses have ranges within a factor of 2 of the noise floor in Fig.  \ref{fig:Noise}. The symbol colors indicate the same classes as above, but are now based on the values of the Pearson-r coefficients computed here. For example, the red points have positive slopes and r-coefficients greater than 0.3; there are fewer of them than before. The periods are almost entirely from \citet{McQ14}. They were checked against our own version of an autocorrelation method and there was excellent agreement in almost all cases. A few appeared either significantly too long or much too short. The main difference is that the 6 periods in the sample that are greater than 30 days in \citet{McQ14} are determined to be under 30 days here, with periods typically less than half of theirs. The reasons for these corrections are discussed in more detail in the Appendix.

Four of the longest period facular cases have been shortened to below 20 days (see discussion above). The proposed period changes have an effect on the conclusions about the distribution of stars with positive correlation slopes, as discussed below. It is apparent in Fig. \ref{fig:PerCorrl}  that the relatively rare facular stars are spread throughout the periods, and that there is some tendency for the spot-dominated stars to have steeper slopes in the middle periods. Also shown in yellow are the three instances of the Sun analyzed in Section \ref{sec:Sun}. The highest of these points belongs to the quiet Sun and the lowest to the active Sun. The upper one would be red and the lower two would actually be black (with a Pearson-r values of 0.19 and -0.21).

Some of the MTF17 facular (red) points have moved into the ``uncertain" (black) class; a number (and much larger fraction) of the remaining facular stars have smaller correlation slopes than most of the spot-dominated cases. The result is that the fraction of facular stars at long periods is less striking than in MTF17. Because these changes are perhaps arguable, we also produced a version of Fig. \ref{fig:PerCorrl} using only the \citet{McQ14} periods. The main effect is to slightly lower the red histogram between 15-20 days and redistribute those cases as individual points in the 45-55 day region. The red histogram numbers at periods below 3 days are also slightly increased at the expense of the points just above 3 days. Finally, the histogram of stars whose brightness is not strongly correlated with variability (black line) rises at short periods; most of them have periods of less than 10 days.

MTF17 draw one of their main conclusions from the fact that there is a sharp drop-off in the {\it fractional} spot-dominated (``correlated"; negative slope) population compared with the fractional facular (``anti-correlated"; positive slope) population at periods above 15 days. However, their Fig. 8 only shows the distribution above 10 days, while the sample extends below 3 days. As is apparent in the lower panel of Fig. \ref{fig:PerCorrl}, the increase in the {\it fraction} of facular stars occurs at short periods as well as long periods. In both cases this is entirely due to a decrease in the number of spot-dominated cases; the {\it number} of facular stars is relatively independent of period and is too low to provide much statistical certainty. The spot-dominated sample is far larger, and has a somewhat normal distribution, falling off on both sides of 12 days. MTF17 shy away from the shorter period stars because some might be pulsators, but that is only likely below 3 day periods (and less so given the solar temperatures of this sample). The results don't change significantly if one ignores all stars below 3 day periods. They also worry about contamination by F stars, but the narrow temperature range of the sample also makes that unlikely to be a major effect.

A primary conclusion of MTF17 is that the long period fractional change in spot dominance indicates a change in the dynamo behavior between 15 to 25 day periods. It is true that given a star with a rotation period of over 20 days, there is a better chance that it is facular than given a star at 12 days. That is as it should be, since we know that stars become increasingly facular as they become less active (as reviewed in Section \ref{sec:Sun}). But this particular demonstration of that is not very compelling for two reasons. One is the lack of a substantial population with periods longer than 25 days. The other is that the same effect occurs moving shortward for stars with periods under 10 days, which is rather puzzling. 

There could easily be an implicit selection effect in period for the 463 stars isolated by MTF17, since it is well-known that the nature of magnetic activity has a strong dependence on period. For example, \citet{Bas18} find that there is a relation between period and the ratio of time spent by a star with one versus multiple dips in the light curve per rotation (see Section \ref{sec:Models}). This sample is a subset of the stars of like temperature analyzed by \citet{Bas18} so it is no surprise that these stars cleanly display the same dependence of SDR on stellar rotation period that those authors discovered. They also found that the variability amplitude in the single mode is typically twice that during the multiple mode. Thus, stars that have similar durations in both modes (SDR near zero) are most likely to show different photometric variability at different times, and can more easily generate long-term changes in the absolute flux. \citet{Bas18} found the zone of such approximately equal durations is centered around 12 days for stars of solar temperature.

MTF17 pulled their sample of 463 out of a set of initially more than 3800 stars, picking the ones which showed convincing variations in absolute flux. The original large sample has the same basic period distribution as the spot-dominated sample, falling off on both sides of 12 days. It does have a secondary peak at 18 days, and extends more robustly to about 25 rather than 20 days. In order to probe selection effects a little further, we extracted all the stars in the initial MTF17 sample of 3845 stars that are not in their chosen sample of 463 and are brighter than Kep$_{mag}=13.3$. This provided a control sample of 390 stars that were subjected to the same analysis as the primary sample. The spread of variability in the control sample is somewhat smaller than in the primary sample, but the correlations between variability and brightness in the control sample have the same distribution as the primary sample. Most of the control stars are again spot-dominated, and the set of facular stars is much smaller and not very dependent on period. The slopes of the brightness-variability correlations are similar, and the main difference in the samples is that the stars whose Pearson-r coefficients have absolute values less than 0.3 were 22\% of the primary sample but 44\% of the control sample (as might be expected given the selection criteria of MTF17).

\begin{figure}
    \begin{center}
    \epsscale{1.0}
    \plotone{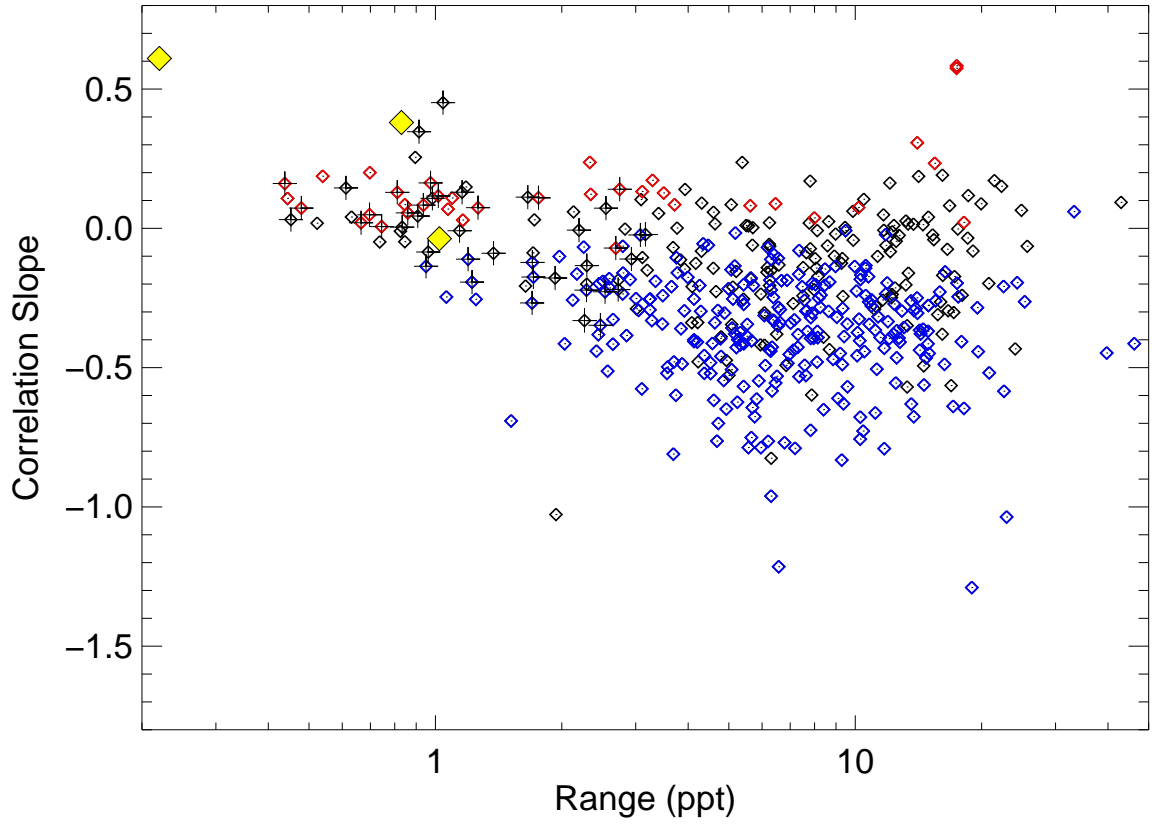}
    \caption{Brightness-Variability correlation slopes versus range ($R_{var}$). The symbols are the same as in Fig. \ref{fig:PerCorrl}. 
    \label{fig:RngCorrl}}
    \end{center}
\end{figure}        

Fig. \ref{fig:RngCorrl} presents an analysis of $R_{var}$ versus correlation slopes. This makes it clear that the stars with larger ranges are preferentially the ones with negative slopes: the spot-dominated cases. There is also a tendency for those with larger negative slopes to have larger values of $R_{var}$. It is more obvious that many of the facular cases are not only quiet, but in the portion of this data set where $R_{var}$ is only 3-5 times greater than the ``noise". The uncorrelated cases occur at all ranges; those uncorrelated at low range seem to be so because of noise issues. Note that the Sun is at the lower end of the observed $R_{var}$ for these Kepler stars, so this is a relatively active sample. The quiet Sun is quieter than any of them; this is likely due to Kepler instrumental noise \citep{Bas13} for the stars. The transitional Sun shows a stronger positive slope than stars with similar $R_{var}$. Still, it is also clear that there are some rather variable stars that show the facular signal (if that is what it is). There are three stars much more variable than the transitional Sun which show a similar positive slope; they have KIC numbers 7221039, 11082584, 11506257 and periods of 4.5, 11.3, and 3.0 days. Their individual point sets all show quite reasonable correlations with strong positive slopes.

It is quite surprising to see any facular stars with short periods. The theory behind the switch bewteen facular and spot domination \citep{Shap14} predicts that short-period stars should be simply spot-dominated, as supported by the fact that they show the highest $R_{var}$ \citep{Bas18}. A closer look at a few of the short-period facular stars reveals they tend to have small values of SDR, and so lie off the general Period-SDR relation. The three that lie on the relation are all individually unconvincing as facular cases when looking at the details of the light curves. Some are stars with rather unchanging $R_{smo}$ which happen to have absolute fluxes a little higher in places where $R_{smo}$ is a little higher. 

Six of them show mixed single/multiple dip modes with what seem like legitimate tendencies for the more variable sections to be brighter in absolute flux. These have KIC numbers 4260947, 6877404, 7221039, 11014223, 11082584, and 11506257. They may require us to re-think the physical reasoning for why their brightness-variability correlations behave as they do. It is possible that faculae are not the reason behind the positive slopes in short-period stars. There is no reason that the solar analogy must apply to these stars. One can imagine extensive micro-flaring on these very active stars as an alternative source of brightening, for example. This mystery remains to be investigated.

\begin{figure}
    \begin{center}
    \epsscale{1.0}
    \plotone{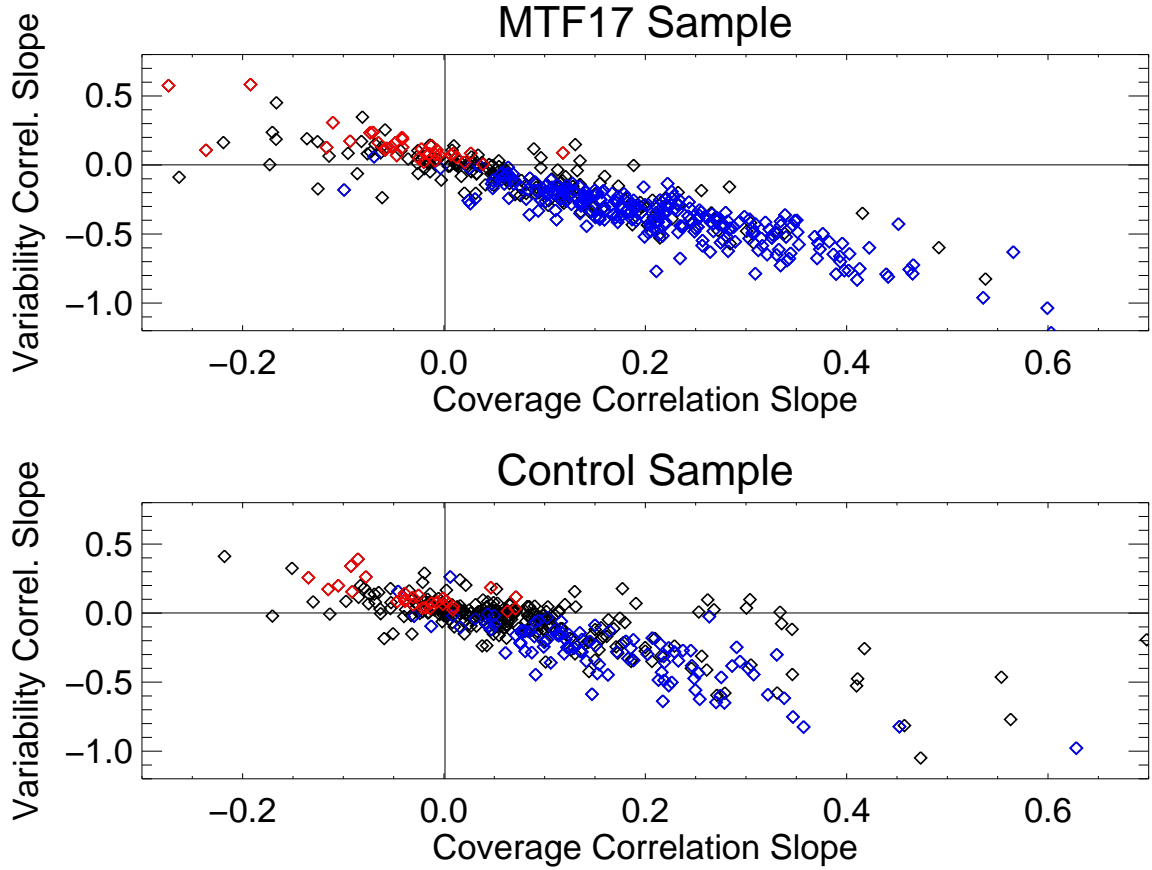}
    \caption{Brightness-Variability correlation slopes compared with Brightness-Coverage correlation slopes. The ordinate in these plots is the same variable as in Fig.  \ref{fig:PerCorrl}. The upper panel shows the results for the primary sample of MTF17, and the lower panel shows the same thing for the control sample. The symbols are the same as in Fig. \ref{fig:RngCorrl}. The variability slopes have been multiplied by 100.
    \label{fig:VarCov}}
    \end{center}
\end{figure}        

Finally, we examine the question of whether spot coverage shares the same correlations with absolute brightness as variability does. If there were no faculae and the measurements of absolute brightness and coverage were as accurate as the measurements of variability are, then one expects that having more coverage would correlate well with lower absolute brightness. Coverage and variability do not have to correlate with each other, since it is possible to have more or less variability without changing the coverage, by putting most of the spots on one side of the star or spreading them more evenly in longitude. Of course, the {\it measurement} of coverage is subject to uncertainties in how to set the unspotted continuum, and the error depends on how the spots are really distributed. 

Figure \ref{fig:VarCov} shows the distributions of the primary and control samples, comparing the correlation slopes for brightness-variability and brightness-coverage with each other (this figure also serves to illustrate the similarity between the results for the two samples). The MidFit solutions were used to compute the coverage. In both samples there is a strong correlation between the two slopes. They almost always have the opposite sign, which fits the naive expectation. That is because the measure of coverage is really the integrated deficit, so it is a smaller number (further below unity) when there are more spots. 

The direct correlations between variability and coverage have almost the same (relatively large) negative values. This is expected because the MidFit solution pulls up regions of lower variability  (implying less coverage). The Flat solutions would produce a much weaker correlation since they tend to have nearly constant coverage independent of variability. There are a few stars in the control sample which have a smaller slope (and correlation coefficient) between brightness and variability, but a greater slope between brightness and coverage. These are all both low confidence correlations and either very short period cases or have major data flaws (large sudden lengthy excursions in the light curve or absolute brightness), so they probably don't tell us anything relevant to this discussion.

\section{Summary and Conclusions}

Several different issues are investigated in this paper. The first is the question of converting Kepler quarterly differential light curves into conjoined light deficit curves suitable for physical starspot analysis. In principle this requires setting the level of the unspotted continuum relative to the differential signal, and removing the effects of quarterly instrumental calibrations. It should also in principle be able to account for the effect of faculae, which increase the apparent brightness of the unspotted star (``unspotted" is different from ``immaculate"; the latter has no magnetic fields present on the surface). One cannot confidently convert the changes in intensity observed in differential light curves into physical information about starspot coverage without solving these issues. 

Despite the superior precision and time coverage of the light curves from these missions, almost all papers concentrate on morphological characteristics of the light curves. One obstacle is the very low spatial resolution inherent in a broad-band (if precise) light curve. A more difficult problem is that there are several mechanisms which can generate similar-looking light curves, but turn out to require different calibration methods for the unspotted continuum. This is made even worse by the independence of the calibration of each quarter of observations in the Kepler dataset. The pipeline reduction scheme does not try to track the absolute brightness of stars, nor is it concerned with an accurate representation of brightness changes on intermediate (one to three month) timescales. A few schemes for overcoming these issues are evaluated, along with trying to understand when they might best be applied. 

The two schemes explored in detail are called the ``Flat" method and the ``MidFit" method. The Flat method uses the whole pipeline differential light curve, which is constructed to have a nearly constant median value over time. It supposes that the highest peaks are closest to the unspotted continuum, which is assumed to be flat (have a constant value relative to the overall differential curve). This sets the inferred level of spot coverage for the rest of the light curve. The less variable segments have lower peaks and smaller dips due to the norming of the pipeline, but represent equally strong spot coverage according to the Flat method of setting the continuum. They are less variable only because spots are inferred to be more evenly spread around the star. The more variable segments are due to a greater concentration of spots on one hemisphere (and thus a relative paucity on the other hemisphere). This is shown to actually be the appropriate interpretation if the changes in variability are primarily due to differential rotation that dominates over changes due to spot evolution. A morphological signature of this condition might be that the light curve will tend to exhibit a single dip in the more variable segments and multiple (usually two) dips in the less variable segments. This will be investigated further in the later modeling paper.

The ``MidFit" method bins the peaks in the light curve over 8 rotation periods and then flattens the remaining variations out by division with a spline fit to these binned points. This turns out to be the most generally useful method, and does a good job of converting many types of differential curves into reasonable facsimiles of the true deficit curves from starspot models. It does worse when there are long enough coherent patches of greater or lesser variability (such as in pure differential rotation models), or when spot evolution significantly changes the total spot coverage for many rotations. Because it sets the continuum ``locally" (over about 10-15 rotations) it will minimize coverage on those timescales and not necessarily register a long period through which the peaks of the differential light curve are well below the true continuum.

The two methods are tested on data from Cycle 23 of the Sun, where actual spot coverage from solar images is available for comparison. The absolute solar light curve is ``Keplerized" (converted to a Kepler-like form) by dividing it into time intervals like quarters, and removing intermediate trends in the medians of each segment to produce a differential light curve, then normalizing all the medians to unity. The MidFit method does a reasonable job of recovering the known spot areal coverage for the active Sun, suffers a bit from the changing coverage in the transitional Sun, and has a problem due to the fact that most variations in the quiet Sun are not caused by spots but rather by faculae. Stars with magnetic activity have not only starspots but also faculae, which are more diffuse magnetic features that increase the stellar brightness (and are more prominent near the limb rather than disk center). Their effect on light curves is more subtle, and they are best identified with absolute rather than relative photometry. Since they are included in the solar signal, we show that the proposed calibration methods are not sensitive to the presence of faculae on the Sun. It is well known that the Sun is brighter overall when it is active compared to when it is quiet, even though the presence of dark sunspots is much greater when it is active. This is shown (and known) to be an over-generalization; on the active Sun spots begin to darken it overall more than faculae brighten it. 

We also utilize selected model light curves that resemble illustrative Kepler observations in the sample from an extensive parameter exploration of spot models, as another test of the merits of the proposed methods for calibrating deficit light curves. Both methods are applied to Keplerized model outputs and the results compared to the correct model output. A few direct comparisons of Kepler observations to qualitatively similar model light curves are also provided. Some light curves can be dominated by differential rotation while others are dominated by spot evolution, but it is currently hard to confidently separate the two. A given star may also switch modes, for example if the latitudinal coverage changes significantly thus making more of the shear visible, or during activity cycles in which the coverage changes substantially and spot evolution becomes more important to the light curve morphology. A fully correct calibration appears to depend on knowing more about this, so we will need to develop diagnostics from differential light curves that better identify which of the proposed methods should be applied to which observational segments. Nonetheless, there are many cases in which the MidFit method works quite well (often along with the Flat method), so it is fair to proceed with some caution. It is particularly favorable to do so when both agree with each other. When they don't, the MidFit method is to be preferred unless there is a good reason to believe that differential rotation is responsible for the major shifts in variability (in which case the Flat method is to be preferred).

We study the very helpful absolute photometry from Kepler FFI images by \citet{Mont17} (MTF17) to determine whether it could help with the determination of the deficit curves (especially by correcting for the effects of faculae). It turns out not to be helpful because the absolute calibration must be smoothed enough that the MidFit method removes the remaining information. Without that level of smoothing, applying the raw absolute photometry to the differential light curves yields results that are clearly implausible in many cases (there is too much noise in the current absolute measurements). It would be inappropriate to apply the Flat method to light curves that have first been adjusted (even smoothly) by the absolute photometry, because the method implicitly assumes no facular signal. 

The absolute brightness versus time records from MTF17 are useful for studying the relative contributions of spots and faculae. They performed such an analysis on a stellar sample with effective temperatures near solar and measured rotation periods (463 stars). They studied the correlations between absolute brightness and short-term differential variability (the latter is largely due to spots). We repeat this analysis in more detail, finding good agreement with MTF17 on which stars are spot- or faculae-dominated. Most of the stars in the sample are more photometrically variable than the active Sun and appear to be spot-dominated, with about a fifth being of uncertain correlation. This is unsurprising given that almost all them are more rapid rotators than the Sun. It is unfortunate that there are not many stars with periods longer than 25 days; several previously suggested slow rotators are not confirmed as such. The faculae-dominated population is too small to provide good statistics, is relatively constant with period, and exhibits rather shallow slopes with lower confidence in the correlations. 

MTF17 discuss the {\it fraction} of stars at a given period in the two populations for periods longer than 10 days, and find that the fraction of facular stars compared to spot-dominated stars rises at longer periods. The sample is relatively active, so there is no contradiction between that conclusion and previous ground-based analyses \citep{Hall09} of spot- and faculae-dominated solar-type stars at the long-period end. A change in dynamo mode is not required to explain it, however, just sufficiently decreasing spot coverage \citep{Shap14} as the period increases. The overall sample here peaks at about a 12 day rotation period, and falls off fairly symmetrically on both sides of that for both the spot- and faculae-dominated samples. That means that the fraction of facular stars also rises to short periods below 10 days. Because of this and the small numbers of facular stars we express some skepticism that new conclusions should be drawn about them. 

We also study a control group of stars that were left out of the primary sample by MTF17 because of their less clear variability in absolute photometry. These 390 bright stars from their original much larger sample of FFI-determined absolute brightness curves are analyzed in all the same ways. The same conclusions are drawn for this control sample. It is incompatible with the theory behind spot versus facular dominance \citep{Shap14} for shorter-period very active stars to show fractionally increasing facular dominance. The apparent facular signal is both weak and only present in a small fraction of stars, and shows no period dependence. We therefore wonder whether the FFI-generated absolute photometry is measuring a cleanly facular signal, particularly at short periods. If the signal holds up, a non-solar explanation must be sought. It will be valuable to extend this type of investigation to new data from K2 and TESS (although they have much shorter observing windows).

Returning to the main point of this paper, a lot of work remains before we can confidently interpret the morphologies of differential light curves in terms of relevant physical variables like spot coverage (and by extension, activity cycles), differential rotation, and spot evolution. An extensive exploration throughout a large physical parameter space of starspot models is needed to make progress toward that goal. A paper in progress studies morphological characteristics of light curves like the single/double dip ratio (SDR), periodogram statistics, intermediate and long-term variability structures, and coverage behavior over time. Hopefully we can identify diagnostics in differential light curves that will allow a better exploitation of the extensive uniquely precise and years-long set of stellar light curves that are a treasure from the Kepler mission which will not be repeated soon. 

\acknowledgments

I am grateful to Ben Montet for helpful discussions and making the data from MTF17 easily available. I would like to acknowledge the contributions of three UC Berkeley undergraduate students. Jner Tzern ``JJ" Oon contributed to the initial analysis of MTF17 results. Riya Shah downloaded and prepared the light curves from the MAST and performed some model computations. Sanah Bhimani contributed to the new analysis of rotation periods. This paper includes data collected by the Kepler mission. Funding for the Kepler mission was provided by the NASA Science Mission directorate. Most of the data presented in this paper were obtained from the Mikulski Archive for Space Telescopes (MAST) at STScI, which is operated by the Association of Universities for Research in Astronomy, Inc., under NASA contract NAS5-26555. I thank the referee for numerous suggestions that improved the clarity of the presentation. 

{\it Facilities:} \facility{Kepler}, \facility{MAST}.

\appendix

\section{Re-evaluating Rotation Periods}

\begin{figure}
    \begin{center}
    \epsscale{1.0}
    \plotone{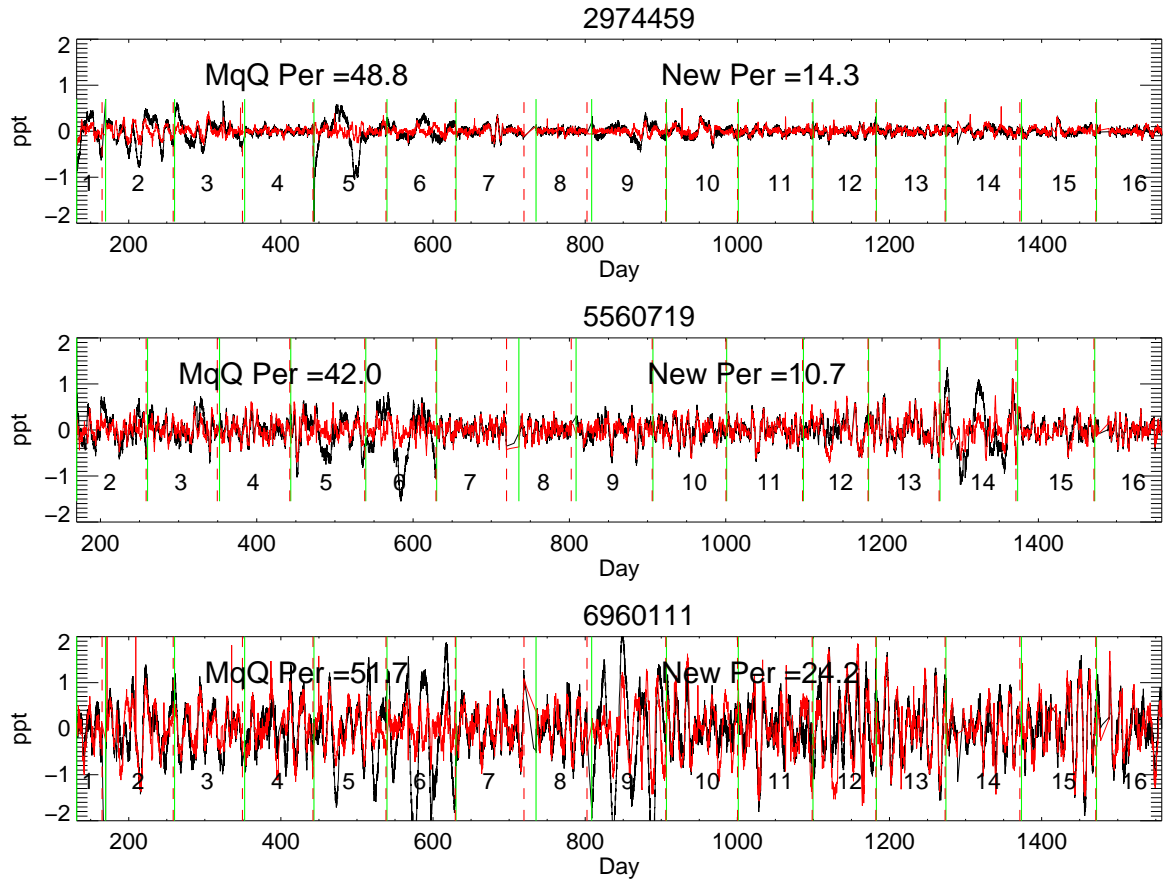}
    \caption{Light Curves for the Questioned Long Period Stars (a). Each panel shows one of the stars, with the old version of the light curve in black and the new version (DR25) in red. The old and new period determinations are also listed, along with the quarter numbers below.
    \label{fig:PerFix1}}
    \end{center}
\end{figure}

\begin{figure}
    \begin{center}
    \epsscale{1.0}
    \plotone{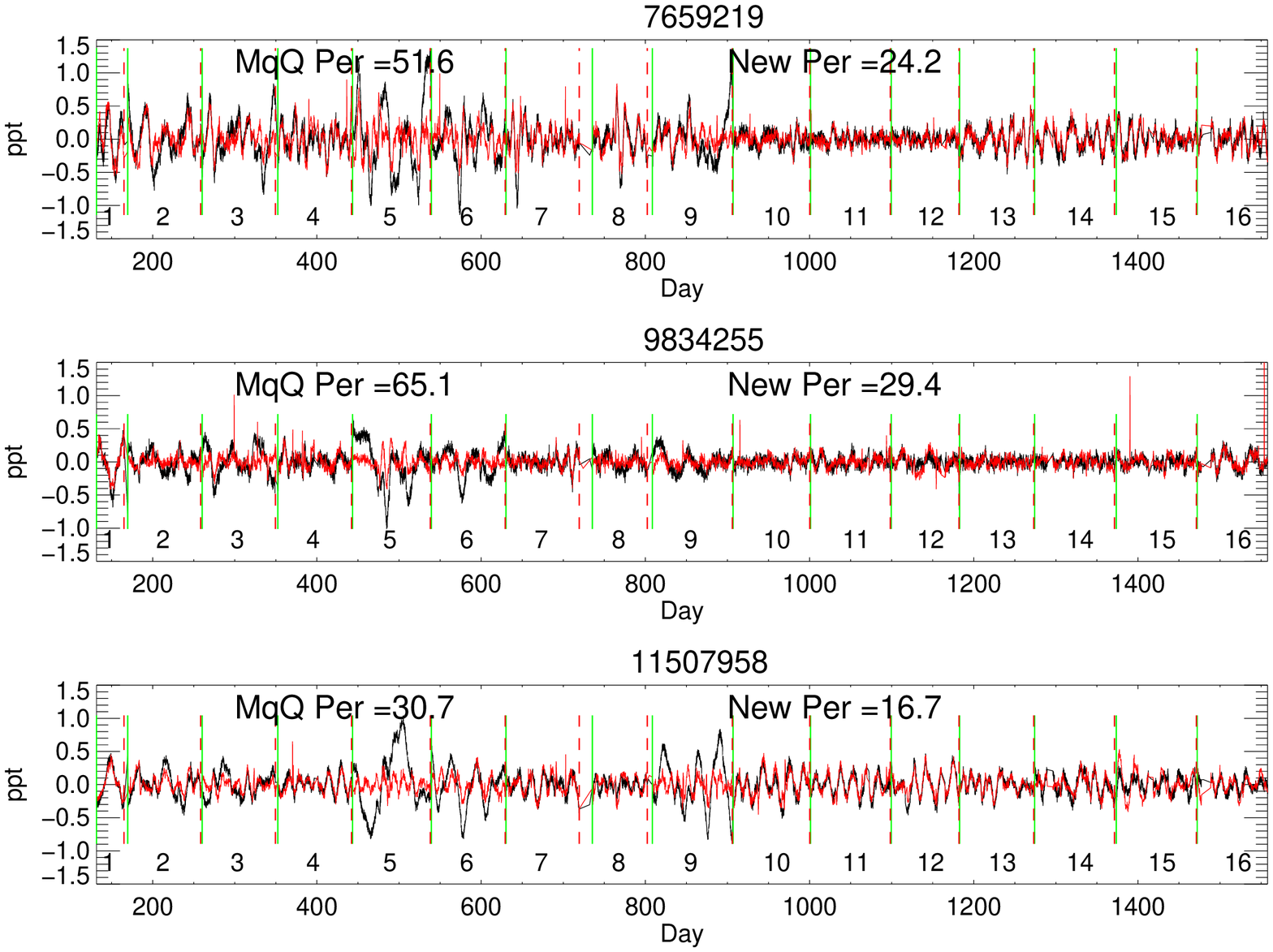}
    \caption{Light Curves for the Questioned Long Period Stars (b). The other three stars whose periods we shorten.
    \label{fig:PerFix2}}
    \end{center}
\end{figure}

Because several periods for the long period stars from \citet{McQ14} are called into question in this paper and shortened, a more explicit justification is called for. The stars in question are shown in Figs. \ref{fig:PerFix1} and \ref{fig:PerFix2}. There are large slow features in certain quarters (5,6, and 9 for all and 14 for 5560719) in the older (black) versions of the light curves that are not present in the DR25 (red) versions. They are suspicious both because they tend to occur in the same quarters and because they are not present in the latest pipeline product. Long-period ($>35$ day) autocorrelation minima are not the first strong minima found in the DR25 versions, but do arise from the apparently spurious features in the earlier versions of the light curves that \citet{McQ14} used (DR14-19). Because they used long stretches of the light curves (Q3-Q14) whereas we correlate quarters two at a time and step through Q2-Q16, we are better able to monitor when autocorrelation minima come and go during the whole observation.

There is often a danger of choosing the half harmonic rather than the true period, particularly given the single/double dip character of the light curves \citep{Bas18}. Each of the DR25 light curves exhibits strong autocorrelation minima at periods shorter than 35 days. In the light curves themselves, oscillations occur regularly 4-6 times per quarter. These could arise if the period were 40-60 days only if the stars are dominated by the double-dip mode. Single dip stars with the original long periods would produce 2-3 oscillations per quarter (as with the spurious features).  Significant autocorrelation minima much shorter than half the adopted period are a warning against that period, as they are hard to produce by rotating spots (it is difficult to make a light curve with 3 significant minima, even with 3 equally spaced spots). For all the DR25 versions of the stars, there are autocorrelation minima that are too closely spaced to be compatible with the original long periods.

\begin{figure}
    \begin{center}
    \epsscale{1.0}
    \plotone{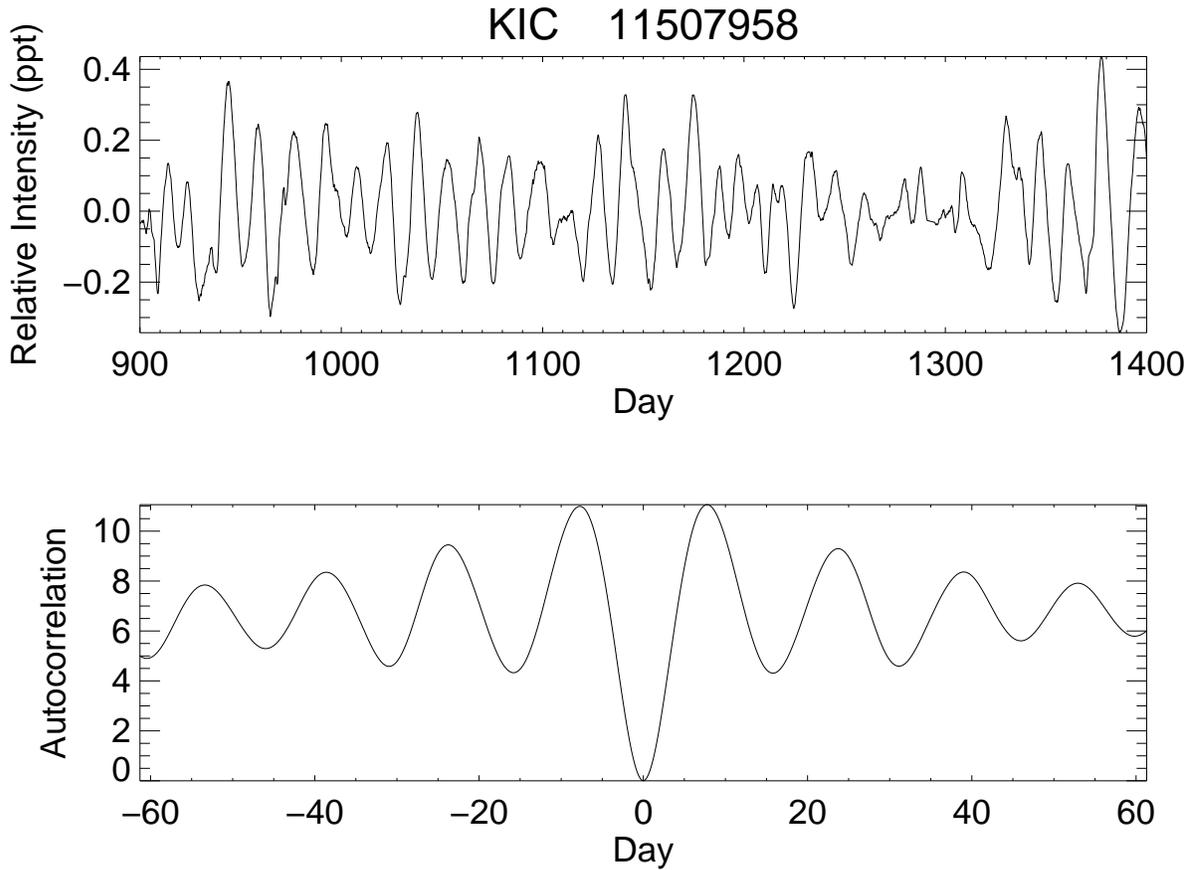}
    \caption{Detail for KIC 11507958. The upper plot shows a portion of the DR25 light curve for this star, and the lower plot shows the autocorrelation function for that segment. The light curve has been smoothed by a 3 day boxcar for clarity. The autocorrelation function shows its first dip at half the rotation period from \citet{McQ14}, and equally spaced dips that get steadily smaller.
    \label{fig:Xcorl}}
    \end{center}
\end{figure}

In a couple of cases we believe they chose the wrong harmonic. One example is shown in Fig. \ref{fig:Xcorl}, where we have cut the original (not overly long) rotation period in half. It shows a detail of the DR25 light curve from the last star in Fig. \ref{fig:PerFix2}, KIC 11507958. In this case the short-period light curve oscillations are large and frequent enough that they dominate the autocorrelation function, and the spurious features do not produce the most significant minima even in the old light curve. The autocorrelation function shows its first dip at half the rotation period given by \citet{McQ14}. Using their (and our) procedure, the period should be assigned to this dip (16.7 days), since the second dip should only be chosen if the first dip is not as pronounced. Apparently that was the case in their analysis, since their period is 30.7 days. Here it is clear that the autocorrelation features get steadily smaller, so the first one should be assigned the rotation period. Examining the light curve itself, it is also clear that there is not a pattern of asymmetry that repeats with a 2-dip cadence, which should be present at least some of the time if this were a double-dip star with a rotation period of about 31 days. Similar reasoning applies to KIC 6960111 (where we have again cut the period in half).

All these stars have low $R_{var}$, which is more compatible with their originally longer periods, and a little puzzling given the new periods. One possibility is that these stars correspond to a type of  ``false-positive" cases for KOIs, in which a brighter quiet foreground star is spatially superposed on a fainter background active (shorter period) star. It is probably time to revisit all the Kepler rotation periods and evaluate the role of noise and spurious instrumental features again. Now that we are aware that the single/double dip character of stars is important and evolves in time, there may be more careful ways of eliminating harmonics. It is also important, however, to close by noting once again that the vast majority of the periods in \citet{McQ14} continue to hold up well (and are the same in our assessment).

\end{document}